\documentstyle[12pt]{article}

\newread\epsffilein    % file to \read
\newif\ifepsffileok    % continue looking for the bounding box?
\newif\ifepsfbbfound   % success?
\newif\ifepsfverbose   % report what you're making?
\newdimen\epsfxsize    % horizontal size after scaling
\newdimen\epsfysize    % vertical size after scaling
\newdimen\epsftsize    % horizontal size before scaling
\newdimen\epsfrsize    % vertical size before scaling
\newdimen\epsftmp      % register for arithmetic manipulation
\newdimen\pspoints     % conversion factor
\def\epsfbox#1{\global\def\epsfllx{72}\global\def\epsflly{72}%
   \global\def\epsfurx{540}\global\def\epsfury{720}%
   \def\lbracket{[}\def\testit{#1}\ifx\testit\lbracket
   \let\next=\epsfgetlitbb\else\let\next=\epsfnormal\fi\next{#1}}%
\def\epsfgetlitbb#1#2 #3 #4 #5]#6{\epsfgrab #2 #3 #4 #5 .\\%
   \epsfsetgraph{#6}}%
\def\epsfnormal#1{\epsfgetbb{#1}\epsfsetgraph{#1}}%
\def\epsfgetbb#1{%
\openin\epsffilein=#1
\ifeof\epsffilein\errmessage{I couldn't open #1, will ignore it}\else
   {\epsffileoktrue \chardef\other=12
    \def\do##1{\catcode`##1=\other}\dospecials \catcode`\ =10
    \loop
       \read\epsffilein to \epsffileline
       \ifeof\epsffilein\epsffileokfalse\else
          \expandafter\epsfaux\epsffileline:. \\%
       \fi
   \ifepsffileok\repeat
   \ifepsfbbfound\else
   \ifepsfverbose\message{No bounding box comment in #1; using defaults}\fi\fi
   }\closein\epsffilein\fi}%
\def\epsfsetgraph#1{%
   \epsfrsize=\epsfury\pspoints
   \advance\epsfrsize by-\epsflly\pspoints
   \epsftsize=\epsfurx\pspoints
   \advance\epsftsize by-\epsfllx\pspoints
   \epsfxsize\epsfsize\epsftsize\epsfrsize
   \ifnum\epsfxsize=0 \ifnum\epsfysize=0
      \epsfxsize=\epsftsize \epsfysize=\epsfrsize
     \else\epsftmp=\epsftsize \divide\epsftmp\epsfrsize
       \epsfxsize=\epsfysize \multiply\epsfxsize\epsftmp
       \multiply\epsftmp\epsfrsize \advance\epsftsize-\epsftmp
       \epsftmp=\epsfysize
       \loop \advance\epsftsize\epsftsize \divide\epsftmp 2
       \ifnum\epsftmp>0
          \ifnum\epsftsize<\epsfrsize\else
             \advance\epsftsize-\epsfrsize \advance\epsfxsize\epsftmp \fi
       \repeat
     \fi
   \else\epsftmp=\epsfrsize \divide\epsftmp\epsftsize
     \epsfysize=\epsfxsize \multiply\epsfysize\epsftmp   
     \multiply\epsftmp\epsftsize \advance\epsfrsize-\epsftmp
     \epsftmp=\epsfxsize
     \loop \advance\epsfrsize\epsfrsize \divide\epsftmp 2
     \ifnum\epsftmp>0
        \ifnum\epsfrsize<\epsftsize\else
           \advance\epsfrsize-\epsftsize \advance\epsfysize\epsftmp \fi
     \repeat     
   \fi
   \ifepsfverbose\message{#1: width=\the\epsfxsize, height=\the\epsfysize}\fi
   \epsftmp=10\epsfxsize \divide\epsftmp\pspoints
   \vbox to\epsfysize{\vfil\hbox to\epsfxsize{%
      \includegraphics{#1}%
      \hfil}}%
\epsfxsize=0pt\epsfysize=0pt}%

{\catcode`\%=12 \global\let\epsfpercent=%\global\def\epsfbblit{%BoundingBox}}%
\long\def\epsfaux#1#2:#3\\{\ifx#1\epsfpercent
   \def\testit{#2}\ifx\testit\epsfbblit
      \epsfgrab #3 . . . \\%
      \epsffileokfalse
      \global\epsfbbfoundtrue
   \fi\else\ifx#1\par\else\epsffileokfalse\fi\fi}%
\def\epsfgrab #1 #2 #3 #4 #5\\{%
   \global\def\epsfllx{#1}\ifx\epsfllx\empty
      \epsfgrab #2 #3 #4 #5 .\\\else
   \global\def\epsflly{#2}%
   \global\def\epsfurx{#3}\global\def\epsfury{#4}\fi}%
\def\epsfsize#1#2{\epsfxsize}
\newcount\figno
\figno=0
\def\fig#1#2#3{
\par\begingroup\parindent=0pt\leftskip=1cm\rightskip=1cm\parindent=0pt
\baselineskip=11pt
\global\advance\figno by 1
\vskip 12pt
\epsfxsize=#3
\centerline{\raisebox{5truecm}{${\displaystyle
{c:\updownarrow\atop h:\leftrightarrow}}$}$\!\!\!\!\!\!\!\!$\epsfbox{#2}}
\vskip 12pt
\begin{quote}{\bf Figure \the\figno:} #1\end{quote}\par
\endgroup\par}
\def\figlabel#1{\xdef#1{\the\figno}}
\def\mulfig#1#2#3#4{
\par\begingroup\parindent=0pt\leftskip=1cm\rightskip=1cm\parindent=0pt
\baselineskip=11pt
\global\advance\figno by 1
\vskip 12pt
\centerline{\epsfxsize=#4\epsfbox{#2}\epsfxsize=#4\epsfbox{#3}}
\vskip 12pt
\begin{quote}{\bf Figure \the\figno:} #1\end{quote}\par
\endgroup\par}
\makeatletter
\ifcase\@ptsize
  \font\tenmsy=msbm10
  \font\sevenmsy=msbm7
  \font\fivemsy=msbm5
  \font\tenmsx=msam10
  \font\sevenmsx=msam7
  \font\fivemsx=msbm5
\or
  \font\tenmsy=msbm10 scaled \magstephalf
  \font\sevenmsy=msbm8
  \font\fivemsy=msbm6
  \font\tenmsx=msam10 scaled \magstephalf
  \font\sevenmsx=msam8
  \font\fivemsx=msam6
\or
  \font\tenmsy=msbm10 scaled \magstep1
  \font\sevenmsy=msbm8
  \font\fivemsy=msbm6
  \font\tenmsx=msam10 scaled \magstep1
  \font\sevenmsx=msam8
  \font\fivemsx=msam6
\fi
\newfam\msyfam
\textfont\msyfam=\tenmsy  
\scriptfont\msyfam=\sevenmsy
\scriptscriptfont\msyfam=\fivemsy
\newfam\msxfam
\textfont\msxfam=\tenmsx  
\scriptfont\msxfam=\sevenmsx
\scriptscriptfont\msxfam=\fivemsx
\def\Bbb{\ifmmode\let\next\Bbb@\else
\def\next{\errmessage{Use \string\Bbb\space only in math mode}}\fi\next}
\def\Bbb@#1{{\Bbb@@{#1}}}
\def\Bbb@@#1{\fam\msyfam#1}
\def\Aaa{\ifmmode\let\next\Aaa@\else
\def\next{\errmessage{Use \string\Aaa\space only in math mode}}\fi\next}
\def\Aaa@#1{{\Aaa@@{#1}}}
\def\Aaa@@#1{\fam\msxfam#1}
\newfam\euffam
\font\sixeuf=eufm6
\font\eighteuf=eufm8
\font\twelveeuf=eufm10 scaled\magstep1
\textfont\euffam=\twelveeuf
\scriptfont\euffam=\eighteuf
\scriptscriptfont\euffam=\sixeuf

\makeatother
\newcommand{\BC}{{\Bbb{C}}}

\newcommand{\BQ}{{\Bbb{Q}}}

\newcommand{\BZ}{{\Bbb{Z}}}
\def\pn{\par\noindent}
\def\nn{\noindent}

\def\W{${\cal W}$}
\def\vec#1{{\rm\bf#1}}
\def\vac#1{|{#1}\rangle}
\def\Vac#1{\left|{#1}\right\rangle}
\def\avac#1{\langle{#1}|}
\def\Avac#1{\left\langle{#1}\right|}

\def\vev#1#2{\langle{#1}|{#2}\rangle}

\def\w{{\cal W}}
\def\del{\partial}

\def\be{\begin{equation}}
\def\ee{\end{equation}}
\def\ba{\begin{array}}
\def\ea{\end{array}}
\def\bea{\begin{eqnarray}}
\def\eea{\end{eqnarray}}
\def\bean{\begin{eqnarray*}}
\def\eean{\end{eqnarray*}}
\def\bl{\begin{list}{}{}}

\def\ts{\textstyle}
\def\fr#1#2{{\ts\frac{#1}{#2}}}

\def\ph{\phantom{--}}

\renewcommand{\theequation}{\mbox{\arabic{section}.\arabic{equation}}}
\newcommand{\reseteqn}{\setcounter{equation}{0}}
\newcommand{\mysection}{\reseteqn\section}

\if@twoside
   \oddsidemargin 0.5cm
   \evensidemargin 0cm
   \marginparwidth 0pt
\else
   \oddsidemargin 0.5cm
   \evensidemargin -0.7cm
   \marginparwidth 0pt
\fi
\marginparsep 0pt
\topmargin 0cm
\headheight 0pt
\headsep 0pt
\topskip 1pt
\footheight 12pt
\footskip 30pt
\textwidth 16.5cm
\textheight 22cm
\columnsep 10pt
\columnseprule 0pt

\begin{document}
  \pagestyle{empty}
  \begin{raggedleft}
IASSNS-HEP-97/73\\
hep-th/9707090\\
June 1997\\
  \end{raggedleft}
%  \begin{center}
%{\eurm DRAFT}
%  \end{center}
  $\phantom{x}$ %%% \vskip 0.618cm\par
  {\LARGE\bf
  \begin{center}
Singular Vectors\\ 
in\\ 
Logarithmic Conformal Field Theories 
  \end{center}
  }\par
  \vfill
  \begin{center}
$\phantom{X}$\\
{Mic$\hbar$ael A.I.~Flohr\footnote[1]{email: {\tt flohr@sns.ias.edu}}}\\
$\phantom{X}$\\
{\em School of Natural Sciences\\
Institute for Advanced Study\\
Olden Lane\\
Princeton, NJ 08540, USA}
  \end{center}\par
  \vfill
  \begin{abstract}
  \noindent 
Null vectors are generalized to the case of indecomposable representations 
which are one of the main features of logarithmic conformal field theories.
This is done by developing a compact formalism with the particular advantage
that the stress energy tensor acting on Jordan cells of primary fields and 
their logarithmic partners can still be represented in form of linear 
differential operators. Since the existence of singular vectors is subject 
to much stronger constraints than in regular conformal field theory, they 
also provide a powerful tool for the classification of logarithmic conformal 
field theories. 
  \end{abstract}
  \vfill\vfill\vfill
  \newpage
%
%%< INTRODUCTION >%%%%%%%%%%%%%%%%%%%%%%%%%%%%%%%%%%%%%%%%%%%%%%%%%%%%%%%%%
%
  \setcounter{page}{1}
  \pagestyle{plain}
  \mysection{Introduction}

\nn It is now nearly fifteen years ago since the concept of rationality
of conformal field theory (CFT) made its first appearance through the minimal
models of Belavin, Polyakov and Zamolodchikov \cite{BPZ83}. Since then
rational conformal field theories (RCFTs) established themselves as
a main tool in modern theoretical physics.
  
Now it becomes increasingly apparent that so-called logarithmic conformal
field theories (LCFTs), first encountered and shown to be consistent in
\cite{Gur93}, are not just a peculiarity but merely a generalization of
ordinary 2-dimensional CFTs with broad and growing applications. 
One may well say that LCFTs contain ordinary rational conformal field
theories (RCFTs) as just the subset of theories free of logarithmic
correlation functions. However, logarithmic divergences are sometimes
quite physical, and so there is an increasing interest in these logarithmic
conformal field theories. 

These logarithms have been found by now in a multitude of models such as the
WZNW model on the supergroup $GL(1,1)$ \cite{RoSa92}, the $c_{p,1}$ models 
(as well as non-minimal $c_{p,q}$ models)
\cite{Flo95,GaKa96,Gur93,Kau91,Kau95,Roh96}, gravitationally dressed conformal
field theories \cite{BiKo95}, WZNW models at level $0$ \cite{KoMa95,CKLT97},
and some critical disordered models \cite{CKT95,MaSe96}. Also, the Yangian
structure of WZNW models seems to be connected to LCFTs \cite{BMM96}.
The theory of indecomposable representations of the Virasoro algebra (which 
are a particular feature of LCFTs in general) was developed in \cite{Roh96}, 
and logarithmic correlation functions were considered in general in
\cite{GhKa97,KoLe97,KLS97,RAK96,ShRa96}, see also \cite{RaRo97} about
consequences for Zamolodchikov's $C$-theorem.

First applications to physical systems include the study of (multi-)critical 
polymers and percolation in two dimensions 
\cite{DuSa87,Flo95,Sal92,Wat96}, two-dimensional turbulence and
magneto-hydrodynamics \cite{Flo96a,RaRo95}, the quantum Hall effect
\cite{Flo96,GFN97,MaSe96,WeWu94}, and also gravitational dressing 
\cite{BiKo95,KoLe97,KLS97} as well as disorder and localization effects
\cite{CKT95,KMT96,MaSe96}. 
They also play a role in the so called unifying $\w$ algebras \cite{BEHHH94}
and are believed to be important for studying the problem of recoil
in the theory of strings  and $D$-branes
\cite{BCFPR96,EMN96,KoMa95,KMW96,PeTa96} as well as target-space 
symmetries in string theory in general \cite{KoMa95}. 

Although LCFTs are mainly considered with respect to the Virasoro algebra,
the concept is more general allowing for Jordan cell structures with
respect to extended chiral symmetry algebras (e.g.\ current algebras) 
as first introduced in \cite{KLS97}. Let us briefly recall what We mean by
Jordan cell structure. Suppose we have two operators $\Phi(z),\Psi(z)$ with
the same conformal weight $h$. As was first realized in \cite{Gur93}, this
situation leads to logarithmic correlation functions and to the fact that
$L_0$, the zero mode of the Virasoro algebra, can no longer be diagonalized:
  \bea
  L_0\vac{\Phi} & = & h\vac{\Phi}\,,\nonumber\\
  L_0\vac{\Psi} & = & h\vac{\Psi} + \vac{\Phi}\,,
  \eea
where we worked with states instead of the fields themselves. The field
$\Phi(z)$ is then an ordinary primary field, whereas the field $\Psi(z)$ 
gives rise to logarithmic correlation functions and is therefore called
a {\em logarithmic partner\/} of the primary field $\Phi(z)$. We would like
to note that two fields of the same conformal dimension {\em do not
not automatically\/} lead to LCFTs with respect to the Virasoro algebra. 
Either, they differ in some other quantum numbers (for examples of such CFTs
see \cite{Flo93,PhD}), or they form a Jordan cell structure with respect to 
an extended chiral symmetry only (see \cite{KoLe97} for a description of
the different possible cases). 

This paper aims in generalizing the concept of singular vectors to the case
of LCFTs, and concentrates -- for the sake of simplicity -- on the case of
LCFTs with respect to the Virasoro algebra. A singular or null vector
$\vac{\chi}$ is a state which is orthogonal to all states,
  \be
  \vev{\psi}{\chi} = 0\ \forall \vec{\psi}\,,
  \ee
where in our case the scalar product is given by the Shapovalov form. Such
states can be considered to be identically zero, and it is precisely the
existence of such states which ``makes'' CFTs (quasi-)rational by dividing
the ideals generated by them out of the Verma modules. This is, of course, well 
known since \cite{BPZ83,FeFu83}, leading to degenerate conformal families etc. 

A pair of fields $\Phi(z),\Psi(z)$ forming a Jordan cell structure brings
the problem of off-diagonal terms produced by the action of the Virasoro
field, such that the corresponding representation is indecomposable. 
Therefore, if $\Vac{\chi^{}_{\Phi}}$ is a null vector in the Verma
module on the highest weight state $\vac{\Phi}$ of the primary field, we 
cannot just replace $\vac{\Phi}$ by $\vac{\Psi}$ and obtain another null 
vector. 

Before we define general null vectors for Jordan cell structures, we
present a formalism which might be useful in the future for all kinds of
explicit calculations in the LCFT setting. This formalism, which we introduce
in section two, has the advantage that the Virasoro modes are still
represented as linear differential operators, and that it is compact and
elegant allowing for arbitrary rank Jordan cell structures. Moreover, 
the connection between LCFTs and supersymmetric CFTs, which one could glimpse 
here and there \cite{Flo95,CKLT97,RoSa92,Sal92} (see also \cite{EhGa96}), 
seems to be a quite fundamental one. The second half of section two is then 
devoted to apply our formalism to the definition of logarithmic null vectors. 

Section three entirely consists of one very explicit example, our other
explicit results are presented in the Appendix. This example treats the
$c_{3,1}=-7$ model, which is the next model in the series of $c_{p,1}$
LCFTs after the well known $c_{2,1}=-2$ theory. One Jordan cell is spanned
by two fields of conformal weight $h=-\frac{1}{4}$. The primary field 
alone corresponds to the irreducible sub-representation with a null vector
at level two divided out, but the complete Jordan cell should have a
logarithmic null vector at level 4, which is then explicitly constructed.

Next, we consider the consequences of the existence of logarithmic null 
vectors in section four: Their existence is subject to much stronger
constraints as the ordinary null vectors on primary fields are. However,
they do exist also in CFTs with central charge $c=c_{p,q}$ from the
minimal series, if the minimal models are augmented by including certain
fields from outside the conformal grid, as first argued in \cite{Flo95}.
As far as LCFTs with respect to the Virasoro algebra are concerned, we
achieve a classification of all possible cases: Perhaps surprisingly,
only models from the generalized minimal series 
  \be
  c_{p,q} = 1 - 6\frac{(p-q)^2}{pq}\,\ p,q\in\BZ-\{0\}\ {\rm coprime}
  \ee
can feature logarithmic null vectors. On one side, these include rational 
models with $c<1$, i.e.\ augmented minimal models and $c_{p,1}$ models,
as well as a certain $c=1$ model with the conformal weights determined by the 
general formula
  \be
  h_{r,s}(c_{p,q}) = \frac{(pr-qs)^2-(p-q)^2}{4pq}\,\ r,s\in\BZ_+
  \ee
with $p\!=\!q\!=\!1$ (this corresponds to the Gaussian $c=1$ model
at the self-dual radius $R=1/\sqrt{2}$). On the other side, we also find
LCFTs for $c\geq 25$ which formally amounts to replacing $p\mapsto -p$.
These theories are certainly not rational with respect to the Virasoro
algebra alone, nor are they unitary, but may be insofar interesting as this 
is the realm of Liouville
theory with its puncture operator \cite{BiKo95,EMN96,KoLe97}.
We conclude this section with two plots of CFT spectra in the $(h,c)$ plane,
one showing the spectra of ordinary primary fields, the other showing the
spectra of Jordan cells of primary fields and their logarithmic partners. They
nicely visualize arguments of our earlier work (second reference of
\cite{Flo95}) on the origin of LCFTs as limiting points in the space of
RCFTs.

%
%%< SECTION 2 >%%%%%%%%%%%%%%%%%%%%%%%%%%%%%%%%%%%%%%%%%%%%%%%%%%%%%%%%%%%%
%
  \mysection{Setup of Problem and Formalism}

\nn LCFTs are characterized by the fact that some of their highest weight
representations are indecomposable. This is usually described by saying that
two (or more) highest weight states with the same highest weight span 
a non-trivial Jordan cell. In the following we call the dimension of
such a Jordan cell the {\em rank\/} of the indecomposable representation.

Therefore, let us assume that a given LCFT has an indecomposable representation
of rank $r$ with respect to its maximally extended chiral symmetry algebra
$\w$. This Jordan cell is spanned by $r$ states $\vac{w_0,w_1,\ldots;n}$,
$n=0,\ldots,r-1$ such that the modes of the generators of the chiral
symmetry algebra act as
  \bea
  \label{eq:jordan} 
  \Phi^{(i)}_0\vac{w_0,w_1,\ldots;n} &=& w_i\vac{w_0,w_1,\ldots;n}
  + \sum_{k=0}^{n-1}a_{i,k}\vac{w_0,w_1,\ldots;k}\,,\\
  \Phi^{(i)}_m\vac{w_0,w_1,\ldots;n} &=& 0\ {\rm for}\ m>0\,,
  \eea
where usually $\Phi^{(0)}(z) = T(z)$ is the stress energy tensor which gives
rise to the Virasoro field, i.e.\ $\Phi^{(0)}_0 = L_0$, and $w_0 = h$ is the
conformal weight. For the sake of simplicity, we concentrate in this paper
on the representation theory of LCFTs with respect to the pure Virasoro
algbra such that (\ref{eq:jordan}) reduces to
  \bea
  \label{eq:L0}
  L_0\vac{h;n} &=& h\vac{h;n} + (1 - \delta_{n,0})\vac{h;n-1}\,,\\
  L_m\vac{h;n} &=& 0\ {\rm for}\ m>0\,,
  \eea
where we have normalized the off-diagonal contribution to 1. As in ordinary
CFTs, we have an isomorphism between states and fields. Thus, the state
$\vac{h;0}$, which is the highest weight state of the irreducible 
subrepresentation contained in every Jordan cell, corresponds to an ordinary
primary field $\Psi_{(h;0)}(z)\equiv\Phi_{h}(z)$, thereas states $\vac{h;n}$ 
with $n>0$ correspond to the so-called logarithmic partners $\Psi_{(h;n)}(z)$
of the primary field. The action of the modes of the Virasoro field on these 
primary fields and their logarithmic partners is given by
  \be
  \label{eq:vir}
  \Lambda_{-k}(z)\Psi_{(h;n)}(w) = \frac{(1-k)h}{(z-w)^k}\Psi_{(h;n)}(w)
  - \frac{1}{(z-w)^{k-1}}\frac{\del}{\del w}\Psi_{(h;n)}(w)
  - (1 - \delta_{n,0})\frac{\lambda(1-k)}{(z-w)^k}\Psi_{(h;n-1)}(w)\,,
  \ee
with $\lambda$ normalized to 1 in the follwing. As it stands, the off-diagonal
term spoils writing the modes $\Lambda_{-k}(z)$ as linear differential 
operators.

The aim of this section is mainly to prepare a formalism in which the 
Virasoro modes are expressed as linear differential operators. To this end, we
introduce a new -- up to now purely formal -- variable $\theta$ with the
property $\theta^r = 0$. We may then view an arbitrary state in the Jordan
cell, i.e.\ a particular linear combination 
  \be
  \Psi_{h}(\vec{a})(z) = \sum_{n=0}^{r-1}a_n\Psi_{(h;n)}(z)\,,
  \ee
as a formal series expansion describing an arbitrary function $a(\theta)$ 
in $\theta$, namely
  \be
  \label{eq:expand}
  \Psi_h(a(\theta))(z) =
  \sum_{n}a_n\frac{\theta^n}{n!}\Psi_h(z)\,.
  \ee
This means that the space of all states in a Jordan cell can be described
by tensoring the primary state with the space of power series in $\theta$,
i.e. $\Theta_r(\Psi_h)\equiv\Psi_h(z)\otimes\BC[\![\theta]\!]/{\cal I}$, 
where we devided out the ideal generated by the relation ${\cal I}=\langle
\theta^r\!=\!0\rangle$. 
In fact, the action of the Virasoro algebra is now simply given by
  \be
  \label{eq:vir2}
  \Lambda_{-k}(z)\Psi_h(a(\theta))(w) = \left(\frac{(1-k)h}{(z-w)^k}
  - \frac{1}{(z-w)^{k-1}}\frac{\del}{\del w}
  - \frac{\lambda(1-k)}{(z-w)^k}\frac{\del}{\del\theta}\right)
  \Psi_h(a(\theta))(w)\,.
  \ee
Clearly, $\Psi_{(h;n)}(z) = \Psi_h(\theta^n/n!)(z)$, but we will often simplify
notation and just write $\Psi_h(\theta)(z)$ for a generic element in
$\Theta_r(\Psi_h)$. However, the context should always make it clear, whether
we mean a generic element or really $\Psi_{(h;1)}(z)$. The corresponding
states are denoted by $\vac{h;a(\theta)}$ or simply $\vac{h;\theta}$.
To project onto the $k^{{\rm th}}$ highest weight state of the Jordan cell,
we just use
$a_k\vac{h;k} = \left.\del_{\theta}^k\vac{h;a(\theta)}\right|_{\theta=0}$.
In order to avoid confusion with $\vac{h;1}$ we write $\vac{h;\Bbb{I}}$ if
the function $a(\theta)\equiv 1$.

It has become apparent by now that LCFTs are somehow closely linked to
supersymmetric CFTs \cite{Flo95,CKLT97,RoSa92,Sal92} (see also \cite{EhGa96}). 
We suggestively denoted our formal variable by $\theta$,
since it can easily be constructed with the help of Grassmannian variables
as they appear in supersymmetry. Taking $N\!=\!r-1$ supersymmetry with 
Grassmann variables $\theta_i$ subject to $\theta_i^2=0$, we may define
$\theta = \sum_{i=1}^{r-1}\theta_i$. More generally, $\theta$ and its powers
constitute a basis of the totally symmetric, homogenous polynomials in the
Grassmannians $\theta_i$. 

Now we would like to recover $n$-point correlation functions within our
formalism. To this end we consider the general operator product expansion
(OPE) of fields with Jordan cell structure:
  \be
  \Psi_{h_i}(a(\theta))(z)\Psi_{h_j}(b(\theta'))(z') =
  \sum_k a(\theta)b(\theta'){\cal C}_{i,j}^k(z-z')\Phi_{h_k}(z')
  \ee
for $|z-z'|\ll 1$. As in the ordinary case, the homogenous functions turn out
to be ${\cal C}_{i,j}^k(z-z') = C_{i,j}^k(z-z')^{h_k-h_i-h_j}$. But we also
have to expand the product $a(\theta)b(\theta')$ for $\theta-\theta'\rightarrow
0$ such that the right hand side can be rewritten as a sum over certain fields 
$\Psi_{h_k}(f(\theta'))(z')$. In fact, this is easily accomplished in our
formalism by the simple expansion $\theta = \theta' +
\frac{\del}{\del(h_h-h_i-h_j)}$ so that the complete OPE just reads
  \be
  \Psi_{h_i}(a(\theta))(z)\Psi_{h_j}(b(\theta'))(z') =
  \sum_k \Phi_{h_k}(z')\,\left(a(\theta'+\del_{h_k-h_i-h_j})b(\theta')\right)
  C_{i,j}^k(z-z')^{h_k-h_i-h_j}\,.
  \ee
Therefore, the fields on the right hand side are operators of the form
  \be
  \Psi_{h_k}(f(\theta'))(z') = \sum_l\Psi_{h_k}(f_l(\theta'))(z')
  \frac{\del^l}{\del(h_k-h_i-h_j)^l}\,,
  \ee
where $f_l(\theta')=\left.\sum_{m,n,j}\frac{a_m}{m!}\frac{b_n}{n!}
{m\choose j}\theta'^{n+j}\del_h^{m-j}h^l\right|_{h=0}$. From 
this all $n$-point functions can be obtained, if the standard 
$n$-point functions of ordinary primary fields are known. To make the 
expansion of $\theta$ in terms of $\theta'$ plausible, we note that powers
$\theta^m\theta'^n$ would vanish in the limit $\theta\rightarrow\theta'$ 
for $n+m\geq r$ where $r=\min(r_i,r_j)$ with $r_i,r_j$ the ranks of the 
Jordan cell fields of the left hand side. Hence, the product
$a(\theta)b(\theta')$ might become completely zero. However, from general
considerations \cite{GhKa97} we would expect that the fields of the right
hand side have Jordan cell structures with a maximal possible rank 
$r=\max(r_i,r_j)$. This is ensured by an arbitrary ansatz $\theta = \theta' +
{\cal O}$ where ${\cal O}$ can be anything independent of our $\theta$ 
variables. Obviously, ${\cal O}$ cannot be just a $\BC$-term, and it turns out
that ${\cal O}=\del_h$ satisfies all further conditions on a well defined OPE.
The appearance of $\del_h$ will get some additional justification later on.

For completeness, we mention that under conjugation of fields
$\Psi \mapsto \Psi^{\dagger}$, the Jordan cell structure is simply mapped
as $a(\theta)\mapsto a^*(\theta)=\theta^{r-1}a(\theta^{-1})$,
such that e.g.\ the conjugate of the primary field is the top field in the 
Jordan cell. As a consequence, $2$-point functions are only non-zero, if
their degree in $\theta$ is maximal. Since 2-point functions vanish
if the fields are from different Jordan cells, the maximal degree
is the same for both fields and so unambigous. From this follows that all
LCFTs must in particular have the identity in a Jordan cell whose rank
determines the maximal rank of all other fields.\footnote[1]{Here we assumed
that 1-point functions of fields with $h\neq 0$ vanish, which is not 
necessarily true for non-unitary theories \cite{Zam91}. The more general
case of non-zero vacuum expectation values leads to considerable modifications
of all correlation functions, which we will consider in a later work.}
These considerations 
generalize to the case of $n$-point functions. Therefore, we may say that
a LCFT is of rank $r$ if its identity operator forms a Jordan cell of rank
$r$, and it is then convenient to include $\del_{\theta_0}^{r-1}$ into the
measure for correlation functions
  \be 
  \left\langle\prod_{i=1}^n\Psi_{h_i}(a_i(\theta_i))(z_i)\right\rangle =
  \int{\cal D}\Psi\del_{\theta_0}^{r-1}
  \prod_{i=1}^n\Psi_{h_i}(a_i(\theta_i))(z_i)\exp(-S(\Psi))
  \ee
with $\theta_0=\sum_i\theta_i$, such that only correlation functions of
maximal degree in $\theta$ are non-zero ($\theta$ might be any of the 
$\theta_i$ left after contracting the $n$-point function via insertion of
OPEs down to a 1-point function).

Next, we derive the consequences of our formalism. An arbitrary state in a
LCFT of level $n$ is a linear combination of descendants of the form
  \be
  \vac{\psi(\theta)} = \sum_k\sum_{\{n_1+n_2+\ldots+n_m=n\}}
  b^{\{n_1,n_2,\ldots,n_m\}}_k
  L_{-n_m}\ldots L_{-n_2}L_{-n_1}\vac{h;k}
  \ee
which we often abbreviate as
  \be
  \label{eq:descendant}
  \vac{\psi(\theta)} = \sum_{|\vec{n}|=n}
  L_{-\vec{n}}b^{\vec{n}}(\theta)\vac{h}\,.
  \ee
We will mainly be concerned with calculating Shapovalov forms 
$\vev{\psi'(\theta')}{\psi(\theta)}$ which ultimatively cook down (by
commuting Virasoro modes through) to expressions of the form
  \be
  \label{eq:shapovalov}
  \vev{\psi'(\theta')}{\psi(\theta)} = 
  \avac{h';a'(\theta')}\sum_m f_m(c)(L_0)^m\vac{h;a(\theta)}\,,
  \ee
where we explicitly noted the dependence of the coefficients on the central
charge $c$. 
Combining (\ref{eq:shapovalov}) with (\ref{eq:descendant}) we write
$\vev{\psi'(\theta')}{\psi(\theta)}=
\avac{h';a'(\theta')}f_{\vec{n}',\vec{n}}(L_0,C)\vac{h;a(\theta)}$
for the Shapovalov form between two {\em monomial\/} descendants, i.e.\
  \be
  \avac{h';a'(\theta')}f_{\vec{n}',\vec{n}}(L_0,C)\vac{h;a(\theta)} =
  \avac{h';a'(\theta')}L_{n'_1}L_{n'_2}\ldots L_{-n_2}L_{-n_1}\vac{h;a(\theta)}
  \,.
  \ee
More generally, since $L_0\vac{h;a(\theta)} = (h + \del_{\theta})
\vac{h;a(\theta)}$, it is easy to
see that an arbitrary function $f(L_0,C)\in\BC[\![L_0,C]\!]$ acts as
  \be
  \label{eq:fL0}
  f(L_0,C)\vac{h;n} = \sum_k\frac{1}{k!}\left(\frac{\del^k}{\del h^k}f(h,c)
  \right)\vac{h;n-k}\,,
  \ee
and therefore $f(L_0,C)\vac{h;a(\theta)} = \vac{h;\tilde{a}(\theta)}$, where
with $a(\theta) = \sum_n a_n\frac{\theta^n}{n!}$ we have
  \be
  \tilde{a}_n = \sum_k\frac{a_{n+k}}{k!}\frac{\del^k}{\del h^k}f(h,c)\,.
  \ee
This puts the convenient way of expressing the action of $L_0$ on Jordan cells
by derivatives with respect to the conformal weight $h$, which appeared earlier
in the literature, on a firm ground. Moreover, from now on we do not worry 
about the range of summations, since all series automatically truncate in the
right way due to the condition $\theta^r=0$.

It is evident that choosing $a(\theta) = \Bbb{I}$ extracts the irreducible
subrepresentation which is invariant under the action of $L_0$. All other
non-trivial choices of $a(\theta)$ yield states which are not invariant
under the action of $L_0$. The existence of null vectors of level $n$ on such
a particular state is subject to the conditions that
  \be
  \sum_{|\vec{n}|=n}f_{\vec{n}',\vec{n}}(L_0,C) b^{\vec{n}}(\theta,h,c)
  \vac{h} \equiv
  \sum_{|\vec{n}|=n}f_{\vec{n}',\vec{n}}(L_0,C)\sum_k b^{\vec{n}}_k(h,c)
  \vac{h;k} = 0\ \forall\ \vec{n}':|\vec{n}'|=n\,.
  \ee 
Notice that we have the freedom that each highest weight state of the
Jordan cell comes with its own descendants. These conditions determine the 
$b^{\vec{n}}_k(h,c)$ as functions in the conformal
weight and the central charge. Clearly, for $a(\theta) = \Bbb{I}$ this would 
just yield the ordinary results as knonw since BPZ \cite{BPZ83}, 
i.e.\ the solutions for $b^{\vec{n}}_0(h,c)$. 
The question of this paper is, under which circumstances null vectors
exist on the whole Jordan cell, i.e.\ for non-trivial choices of $a(\theta)$.
Obviously, these null vectors, which we call {\em logarithmic null vectors\/}
can only constitute a subset of the ordinary null vectors. From (\ref{eq:fL0})
we immediately learn that the conditions imply
  \be
  \sum_{|\vec{n}|=n}b^{\vec{n}}_k(h,c)
  \frac{\del^k}{\del h^k}f_{\vec{n}',\vec{n}}(h,c) 
  = 0\ \forall\ \vec{n}':|\vec{n}'|=n\,,
  \ee
which can be satisfied if we put 
  \be
  b^{\vec{n}}_k(h,c) = \frac{1}{k!}\frac{\del^k}{\del h^k}b^{\vec{n}}_0(h,c)\,.
  \ee
In fact, choosing the $b^{\vec{n}}_k(h,c)$ in this way allows one to rewrite 
the conditions as total derivatives of the standard condition for
$b^{\vec{n}}_0(h,c)$. Of course, this determines the $b^{\vec{n}}_k(h,c)$
only up to terms of lower order in the derivatives such that the conditions
still take the form 
  \be
  \sum_k\frac{\lambda_k}{k!}\frac{\del^k}{\del h^k}\left(
  \sum_{|\vec{n}|=n}f_{\vec{n}',\vec{n}}(h,c)b^{\vec{n}}_0(h,c)\right)
  = 0\ \forall\ \vec{n}':|\vec{n}'|=n\,,
  \ee
which, however, does not yield any different results. Moreover, the
coefficients $b^{\vec{n}}_k(h,c)$ can only be determined up to an overall
normalization. This means that only $p(n)-1$ of the standard coefficients 
$b^{\vec{n}}_0(h,c)$ are determined to be functions in $h,c$ multiplied by
the remaining coefficient, e.g.\ $b^{\{1,1,\ldots,1\}}_0$ (if this coefficient
is not predetermined to vanish). In order to be able to write the coefficients
$b^{\vec{n}}_k(h,c)$ with $k>0$ as derivatives with respect to $h$, one 
needs to fix the remaining free coefficient $b^{\{1,1,\ldots,1\}}_0=h^{p(n)}$
as a function of $h$. The choice given here ensures that all coefficients
are always of sufficient high degree in $h$.\footnote[1]{Clearly, this works 
only for $h\neq 0$. To find null vectors with $h=0$ needs some extra care. One
foolproof choice is to put the remaining free coefficient to $\exp(h)$, but
we usually choose the least common multiple of the denominators of the
resulting rational functions in $h,c$ of the other coefficients in order to
simplify the calculations. This, however, occasionaly leads to additional --
trivial -- solutions which are the price we way for doing all calculations 
with polynomials only.}

It is important to
understand that this is only a necessary condition due to the
following subtelty: The derivatives with respect to $h$ are done in a purely
formal way. But already determining the standard solution $b^{\vec{n}}_0(h,c)$ 
is not sufficient in itself, and the conditions for the existence of
standard null vectors yield one more constraint, namely $h=h_i(c)$ or vice
versa $c=c_i(h)$ (the index $i$ denotes possible different solutions, since
the resulting equations are higher degree polynomials $\in\BC[h,c]$).
These constraints must be plugged in {\em after\/} performing the
derivatives\footnote[7]{Again, this is not entirely true in the case of $h=0$.
Here, $c=c_i(h)$ has to be plugged in {\em before\/} doing the derivatives.}
and, as it will turn out, this will severely restrict the
existence of logarithmic null vectors, yielding only some {\em discrete\/}
pairs $(h,c)$ for each level $n$. Moreover, the set of solutions gets
rapidly smaller if for a given level $n$ the rank $r$ of the assumed Jordan 
cell is increased. Since there are $p(n)$ linearly independent conditions
for the $b^{\vec{n}}_0(h,c)$ of a standard null vector of level $n$ ($p(n)$
is the number of partitions of $n$), a necessary condition is $r\leq p(n)$.

The next section gives one rather explicit example, but further details
about our calculations (e.g.\ how to find logarithmic null vectors with
$h\!=\!0$) can also be found in the Appendix, where we mainly collect and
comment our explicit results. 

%
%%< SECTION 3 >%%%%%%%%%%%%%%%%%%%%%%%%%%%%%%%%%%%%%%%%%%%%%%%%%%%%%%%%%%%%
%
  \mysection{An Example}

\nn In this section we want to demonstrate what a logarithmic null vector is
and under which conditions it exists. Null vectors are of particular
importance for rational CFTs. For any CFT given by its maximally extended
symmetry algebra $\w$ and a value $c$ for the central chrage we can 
determine the so-called degenerate $\w$-conformal families which contain
at least one null vector. The corresponding heighest weights turn out to
be parametrized by certain integer labels, yielding the so-called
Kac-table. If $\w = \{T(z)\}$ is just the Virasoro algebra, all degenerate
conformal families have highest weights labeled by two integers $r,s$,
  \be
  \label{eq:h}
  h_{r,s}(c) = \frac{1}{4}\left(
  \frac{1}{24}\left(\sqrt{(1-c)}(r+s)-\sqrt{(25-c)}(r-s)\right)^2 
  - \frac{1-c}{6}\right)\,.
  \ee
The level of the (first) null vector contained in the conformal families
over the highest weight state $\vac{h_{r,s}(c)}$ is then $n=rs$.

LCFTs have the special property that there are at least two conformal
families with the same highest weight state, i.e.\ that we must have
$h = h_{r,s}(c) = h_{t,u}(c)$. This does not happen for the so-called minimal
models since their truncated conformal grid precisely excludes this.
However, LCFTs may be constructed for example for $c=c_{p,1}$, where
formally the conformal grid is empty, or by augmenting the field
content of a CFT by considering an enlarged conformal grid. However,
if we have the situation typical for a LCFT, we have two non-trivial and 
{\em different\/} null vectors, one at level $n=rs$ and one at $n'=tu$ where 
we assume without loss of generality $n\leq n'$. Then the null vector at
level $n$ is an ordinary null vector on the highest weight state of the
irreducible sub-representation $\vac{h;0}$ of the rank 2 Jordan cell spanned
by $\vac{h;0}$ and $\vac{h;1}$, but what about the null vector at level $n'$?

Let us consider the particular LCFT with $c=c_{3,1}=-7$. This LCFT admits
the highest weights 
$h\in\{0,\frac{-1}{4},\frac{-1}{3},\frac{5}{12},1,\frac{7}{4}\}$ which yield
the two irreducible representations at $h_{1,3}=\frac{-1}{3}$ and
$h_{1,6}=\frac{5}{12}$ as well as two indecomposable representations with
so-called staggered module structure (roughly a generalization of Jordan cells 
to the case that some highest weights differ by integers \cite{GaKa96,Roh96}) 
constituted by the
triples $(h_{1,1}\!=\!0,h_{1,5}\!=\!0,h_{1,7}\!=\!1)$ and 
$(h_{1,2}\!=\!\frac{-1}{4},h_{1,4}\!=\!\frac{-1}{4},h_{1,8}\!=\!\frac{7}{4})$.
We note that similar to the case
of minimal models we have the identification $h_{1,s} = h_{2,9-s}$ such that
the actual level of the null vector might be reduced. In the following we
will determine the null vectors at level 2 and 4 for the rank 2 Jordan 
cell with $h=\frac{-1}{4}$. First, we start with the level 2 null vector,
whose general ansatz is
  \be
  \vac{\chi^{(2)}_{h,c}} = 
  \left(b^{\{1,1\}}_0L_{-1}^2 + b^{\{2\}}_0L_{-2}\right)
  \vac{h;a(\theta)} +
  \left(b^{\{1,1\}}_1L_{-1}^2 + b^{\{2\}}_1L_{-2}\right)
  \vac{h;\del_{\theta}a(\theta)}\,,
  \ee
where we explicitly made clear how we counteract the off-diagonal action of the 
Virasoro null mode. It is well known that up to an overall normalization we
have
  \be
  b^{\{1,1\}}_0 = 3h\,,\ \ \ \
  b^{\{2\}}_0   = -2h(2h + 1)\,,
  \ee
such that according to the last section we should put
  \be
  b^{\{1,1\}}_1 =  3\,,\ \ \ \
  b^{\{2\}}_1   = -8h - 2\,.
  \ee
The matrix elements $\avac{h}L_2\left.\del_{\theta}^k\vac{\chi^{(2)}_{h,c}}
\right|_{\theta=0}$, $k=0,1$ do give us
further constraints, namely 
  \be
  c = -2h\frac{8h-5}{2h+1}\,,\ \ \ \
  0 = -2h\frac{16h^2+16h-5}{2h+1}\,.
  \ee
{}From these we learn that only for $h\in\{0,\frac{-5}{4},\frac{1}{4}\}$
we may have a logarithmic null vector (with $c=0,25,1$ respectively).
Therefore, the level 2 null vector for $h=\frac{-1}{4}$ of the $c=-7$
LCFT is just an ordinary one.

Next, we look at the level 4 null vector with the general ansatz
  \bea
  \!\!\!\!\!\!\!\!\vac{\chi^{(4)}_{h,c}} &=&
  \left(b^{\{1,1,1,1\}}_0L_{-1}^4 + b^{\{2,1,1\}}_0L_{-2}L_{-1}^2
      + b^{\{3,1\}}_0L_{-3}L_{-1} + b^{\{2,2\}}_0L_{-2}^2
      + b^{\{4\}}_0L_{-4}\right)
  \vac{h;a(\theta)}\\
  &+&
  \left(b^{\{1,1,1,1\}}_1L_{-1}^4 + b^{\{2,1,1\}}_1L_{-2}L_{-1}^2
      + b^{\{3,1\}}_1L_{-3}L_{-1} + b^{\{2,2\}}_1L_{-2}^2
      + b^{\{4\}}_1L_{-4}\right)
  \vac{h;\del_{\theta}a(\theta)}\nonumber\,.
  \eea
Considering the possible matrix elements determines the coefficients up to 
overall normalization as
  \bea
  b^{\{1,1,1,1\}}_0 &=&
  h^4(1232h^3-2466h^2-62h^2c+1198h-296hc+13hc^2+5c^3+92c^2+128c-144)\,,
  \nonumber\\
  b^{\{2,1,1\}}_0   &=&
  -4h^4(1120h^4-2108h^3+140h^3c+428h^2-66h^2c+338h-323hc+90hc^2\nonumber\\
                    & &+\,60c^2-78+99c)\,,\nonumber\\
  b^{\{3,1\}}_0     &=&
  24h^4(96h^5-332h^4+44h^4c+382h^3-8h^3c+4h^3c^2-53h^2c+12h^2c^2-235h^2
  \nonumber\\
                    & &+\,11hc^2+14hc+65h-6+3c+3c^2)\,,\nonumber\\
  b^{\{2,2\}}_0     &=&
  24h^4(32h^3-36h^2+4h^2c+8hc+22h+3c-3)(3h^2+hc-7h+2+c)
  \,,\nonumber\\
  b^{\{4\}}_0       &=&
  -4h^4(550h+3c^3-224h^2c+66hc^2+748h^3-48+2508h^4+11hc^3+41h^2c^2\nonumber\\
                    & &-\,40h^3c-3008h^5+12h^2c^3+120h^3c^2-184h^4c+102hc
                       +27c^2-1698h^2\nonumber\\
                    & &+\,18c+4h^3c^3+768h^6+448h^5c+76h^4c^2)\,.
  \eea
Even for ordinary null vectors at level 4 we have $p(4)=5$ conditions,
but due to the freedom of overall normalization only 4 conditions have
been used so far. The last, $\avac{h}L_{4}\left.\vac{\chi^{(4)}_{h,c}}
\right|_{\theta=0} = 0$, fixes the central charge as a function of the 
conformal weight to 
  \be c\in\left\{
  -2\frac{h(8h-5)}{2h+1},
  -\frac{2}{5}\frac{8h^2+33-41h}{3+2h},
  -\frac{3h^2-7h+2}{h+1},
  1-8h
  \right\}\,.
  \ee
If we again put $b^{\vec{n}}_1(h,c)=\del_h b^{\vec{n}}_0(h,c)$ we obtain the
additional constraint $\avac{h}L_{4}\left.\del_{\theta}^{}
\vac{\chi^{(4)}_{h,c}}\right|_{\theta=0} = 0$, i.e.
  \bea
  \label{eq:loglevel4}
  0 &\!\!=\!\!& 
        -4h^3(-14308h^3c^2+6600h-528c+30hc^3+1239840h^5-113592h^2+5290hc+144c^2
        \nonumber\\
    & & +\,462h^2c^3+4368h^3c^3+275hc^4+360h^2c^4+3296h^4c^3+74240h^6c
        +25632h^5c^2\nonumber\\
    & & +\,67584h^7+595224h^3-25812h^2c-12712h^3c+11574h^2c^2
        -2475hc^2-1287136h^4\nonumber\\
    & & +\,60c^4-249408h^5c+324c^3-12192h^4c^2-504320h^6+187040h^4c+140h^3c^4)
        \,,
  \eea
in which we may insert the four solutions for $c$ to obtain sets of discrete
conformal weights (and central charges in turn). The Appendix contains the
explicit calculations for all possible Virasoro logarithmic null vectors
up to level 5. Here, we are only interested in the null vector for $h=
\frac{-1}{4}$. And indeed, the first two solutions for $c$ admit (among
others) $h=\frac{-1}{4}$ to satisfy (\ref{eq:loglevel4}) with the final
result for the null vector
  \bea
  \Vac{\chi^{(4)}_{h=-1/4,c=-7}} &=&
  \left(\fr{315}{128}L_{-1}^4-\fr{525}{64}L_{-2}L_{-1}^2
  +\fr{315}{128}L_{-2}^2-\fr{105}{64}L_{-3}L_{-1}
  -\fr{105}{64}L_{-4}\right)
  \Vac{\fr{-1}{4};(\alpha_1\theta^1+\alpha_0\theta^0)}\nonumber\\
  &+&\left(-\fr{2463}{128}L_{-1}^4+\fr{2485}{64}L_{-2}L_{-1}^2
  +\fr{1241}{64}L_{-3}L_{-1}-\fr{1383}{128}L_{-2}^2
  +\fr{821}{64}L_{-4}\right)
  \Vac{\fr{-1}{4};(\alpha_1\theta^0)}\,.\nonumber\\
  & &
  \eea
This shows explicitly the existence of a non-trivial logarithmic null vector
in the rank 2 Jordan cell indecomposable representation with highest weight
$h=\frac{-1}{4}$ of the $c_{3,1}=-7$ rational LCFT. Here, $\alpha_0,\alpha_1$
are arbitrary constants such that we may rotate the null vector arbitrarily
within the Jordan cell. However, as long as $\alpha_1\neq 0$, there is 
necessarily always a non-zero component of the logarithmic null vector
which lies in the irreducible sub-representation. Although there is the
ordinary null vector built solely on $\vac{h;0}$, there is therefore no
null vector solely built on $\vac{h;1}$, once more demonstrating the fact
that these representations are indecomposable.

%
%%< SECTION 4 >%%%%%%%%%%%%%%%%%%%%%%%%%%%%%%%%%%%%%%%%%%%%%%%%%%%%%%%%%%%%
%
  \mysection{Kac Determinant and Classification of LCFTs}

\nn As one might imagine from the Appendix, it is quite a time consuming
task to construct logarithmic null vectors explicitly. However, if we are
only interested in the pairs $(h,c)$ of conformal weights and central
charges for which a CFT is logarithmic and owns a logarithmic null vector,
we don't need to work so hard.

As already explained, logarithmic null vectors are subject to the condition
that there exist fields in the theory with identical conformal weights.
As can be seen from (\ref{eq:h}), there are always fields of identical
conformal weights if $c=c_{p,q}=1-6\frac{(p-q)^2}{pq}$ is from the minimal
series with $p>q>1$ coprime integers. However, such fields are to be
identified in these cases due to the existence of BRST charges
\cite{Fel89,FFK89}. Equivalently,
this means that there are no such pairs of fields within the truncated
conformal grid 
  \be\label{eq:confgrid}
  H(p,q) \equiv \{h_{r,s}(c_{p,q}):0<r<|q|,0<s<|p|\}\,. 
  \ee
It is worth noting that our explicit calculations for the data collected in 
the Appendix indeed produced ``solutions'' for the well known null vectors
in minimal models, but these ``solutions'' never had a non-trivial Jordan 
cell structure. For example, at level 3 we find a solution with 
$c=c_{2,5}=-\frac{22}{5}$ and $h=h_{2,1}=h_{3,1}=-\frac{1}{5})$ which, however,
is just the ordinary one. This was to be expected because each Verma module
of a minimal model has precisely two null vectors (this is why all weights
$h$ appear twice in the conformal grid, $h_{r,s}=h_{q-r,p-s}$).
We conclude that logarithmic null vectors can only occur if fields of
equal conformal weight still exist after all possible identifications due
to BRST charges (or due to the embedding structure of the Verma modules
\cite{FeFu83})
have been taken into account. For later convenience, we further define
the boundary of the conformal grid as
  \bea\label{eq:confbound}
  \del H(p,q)   &\equiv& \{h_{r,p}(c_{p,q}):0<r<|q|\} \cup
                         \{h_{q,s}(c_{p,q}):0<s<|p|\} \,,\\
  \del^2 H(p,q) &\equiv& \{h_{q,p}(c_{p,q})\}\,.\nonumber
  \eea
These three sets reflect the possible three embedding structures of the
corresponding Verma modules which are of type $III_{\pm}$, $III_{\pm}^{\circ}$, 
and $III_{\pm}^{\circ\circ}$ respectively \cite{FeFu83}.

In our earlier work we have argued that LCFTs are a very general kind of
conformal theories, containing rational CFTs as the special subclass of
theories without logarithmic fields. In the case of minimal models we
showed that logarithmic versions of a CFT with $c=c_{p,q}$ can be 
obtained by augmenting the conformal grid. This can formally be achieved
by considering the theory with $c=c_{\alpha p,\alpha q}$. The explicit
calculations of null vectors in the present paper, however, did not show
the existence of logarithmic fields for minimal models, the reason being
simply that the levels of null vectors considered here are too small.
Let us look at minimal $c_{2n-1,2}$ models, $n>1$. Fields within the
conformal grid are ordinary primary fields which do not posses logarithmic
partners. Therefore, pairs of primary fields with logarithmic partners have
to be found outside the conformal grid, and according to our earlier work 
\cite{Flo95} and \cite{GaKa96}
must lie on the boundary $\del H(p,q)$ (note that the corner point 
is not an element). Notice that for $c_{p,1}$
models this condition is easily met because the conformal grid $H(p,1)=
\emptyset$. Fields outside the boundary region which have the property
that their conformal weights are $h'=h+k$ with $h\in H(p,q)$, $k\in\BZ_+$ 
do not lead to Jordan cells (they are just descendants of the primary fields). 
For example,
the $c_{5,2}=-\frac{22}{5}$ model admits representations with
$h=h_{1,8}=h_{3,2}=\frac{14}{5}$ which do not form a logarithmic pair and
are just descendants of the $h=-\frac{1}{5}$ representation.
Therefore, even for the $c_{2n-1,2}$ models with their
relatively small conformal grid, the lowest level of a logarithmic null
vector easily can get quite large. In fact, the smallest minimal model, the 
trivial $c_{3,2}=0$ model, can be augmented to a LCFT with formally 
$c=c_{9,6}$ which has 
a Jordan cell representation for $h=h_{2,2}=h_{2,4}=\frac{1}{8}$. 
The logarithmic null vector already has level 8 and reads explicitly 

  {\vskip-0.5ex\footnotesize
  \begin{eqnarray*}
  \lefteqn{\Vac{\chi^{(8)}_{h=1/8,c=0}} =}\\
  & & \left(10800L_{-1}^8-208800L_{-2}L_{-1}^6+928200L_{-2}^2L_{-1}^4
      -1060200L_{-2}^3L_{-1}^2+151875L_{-2}^4+252000L_{-3}L_{-1}^5\right.\\
  & & -\,631200L_{-3}L_{-2}L_{-1}^3+207000L_{-3}L_{-2}^2L_{-1}
      -1033200L_{-3}^2L_{-1}^2+360000L_{-3}^2L_{-2}-1249200L_{-4}L_{-1}^4\\
  & & +\,4165200L_{-4}L_{-2}L_{-1}^2-1133100L_{-4}L_{-2}^2
      +176400L_{-4}L_{-3}L_{-1}+593100L_{-4}^2+624000L_{-5}L_{-1}^3\\
  & & -\,720000L_{-5}L_{-2}L_{-1}-429300L_{-5}L_{-3}+1206000L_{-6}L_{-1}^2
      -455400L_{-6}L_{-2}-206100L_{-7}L_{-1}\\
  & & \left.-\,779400L_{-8}\right)\Vac{\fr{1}{8},a(\theta)}\\
  & \!\!\!\!+\!\!\!\! &
      \left(76800L_{-3}L_{-2}L_{-1}^3+755200L_{-3}L_{-2}^2L_{-1}
      -2596800L_{-3}^2L_{-1}^2+106400L_{-3}^2L_{-2}+179712L_{-4}L_{-1}^4
      \right.\\
  & & +\,123648L_{-4}L_{-2}L_{-1}^2+3621120L_{-4}L_{-3}L_{-1}-857856L_{-4}^2
      +739200L_{-5}L_{-1}^3-5832000L_{-5}L_{-2}L_{-1}\\
  & & \left.+\,992800L_{-5}L_{-3}+3444000L_{-6}L_{-1}^2-154800L_{-6}L_{-2}
      -2210400L_{-7}L_{-1}+488000L_{-8}\right)\Vac{\fr{1}{8},\del a(\theta)}\,,
  \end{eqnarray*}\vskip-0.25ex}

\nn up to an arbitrary state proportional to the ordinary level 4 null vector.
This shows that minimal models can indeed be augmented to logarithmic 
conformal theories. Level 8 is actually the
smallest possible level for logarithmic null vectors of augmented minimal
models. On the other hand, descendants of logarithmic fields are also
logarithmic, giving rise to the more complicated staggered module
structure \cite{Roh96}. Thus, whenever for $c=c_{p,q}$ the conformal weight 
$h=h_{r,s}$ with either $r\equiv 0$ mod $p$, $s\not\equiv 0$ mod $q$, or
$r\not\equiv 0$ mod $p$, $s\equiv 0$, the corresponding representation is
part of a Jordan cell (or a staggered module structure).

The question of whether a CFT is logarithmic really makes sense only in the
framework of (quasi-)rationality. Therefore, we can assume that $c$ and all 
conformal
weights are rational numbers. It can then be shown that the only possible
LCFTs with $c\leq 1$ are the ``minimal'' LCFTs with $c=c_{p,q}$. Using the
correspondence between the Verma modules $V_{h,c}\leftrightarrow
V_{-1-h,26-c}$ one can further show that LCFTs with $c\geq 25$ might exist
with (formally) $c=c_{-p,q}$. Again, due to an analogous (dual) BRST
structure of these models, pairs of primary fields with logarithmic partners
can only be found outside the conformal grid 
$H(-p,q) = \{h_{r,s}(c_{-p,q}):0<r<q,0<s<p\}$, a fact that
can also be observed in our direct calculations. For 
example, at level 4 we found a candidate solution with
$c_{-3,2}=26$ and $h_{4,1}=h_{1,3}=-4$. But again, the explicit calculation
of the null vector did not show any logarithmic part. 

The existence of null vectors can be seen from the Kac determinant of the
Shapovalov form $M^{(n)} = \avac{h}L_{\vec{n}'}L_{-\vec{n}}\vac{h}$,
which factorizes into contributions for each level $n$.
The Kac determinant has the well known form
  \be\label{eq:kac}
  {\rm det}\,M^{(n)} =
  \prod_{k=1}^n\prod_{rs=k}\left(h-h_{r,s}(c)\right)^{p(n-rs)}\,.
  \ee
A consequence of Section 2 is that a necessary condition for the existence
of logarithmic null vectors in rank $r$ Jordan cell representations of
LCFTs is that $\frac{\del^k}{\del h^k}\left({\rm det}\,M^{(n)}\right) = 0$
for $k=0,\ldots,r-1$. It follows immediately from (\ref{eq:kac}) that
non-trivial common zeros of the Kac determinant and its derivatives at level
$n$ only can come from the factors whose powers $p(n-rs)=1$, i.e. $rs=n$ and
$rs=n-1$. For example
  \be\label{eq:dkac}
  \frac{\del}{\del h}\left({\rm det}\,M^{(n)}\right) =
  \sum_{n-1\leq rs\leq n}\frac{1}{\left(h-h_{r,s}(c)\right)}
  \prod_{n-1\leq r's'\leq n}\left(h-h_{r',s'}(c)\right)\,,
  \ee
which indeed yields a non-trivial constraint. Clearly (\ref{eq:dkac}) vanishes
at $h=h_{r,s}(c)$ up-to one term which is zero precisely if there is one other
$h_{t,u}(c)=h$. This is the condition stated earlier. Solving it for the
central charge $c$ we obtain
  \be\label{eq:csol}
  c = \left\{\begin{array}{l}
  {\displaystyle-\frac{(2t-3u+3s-2r)(3t-2u+2s-3r)}{(u-s)(t-r)}}\\
  {\displaystyle-\frac{(2t-3u-3s+2r)(3t-2u-2s+3r)}{(u+s)(t+r)}}
  \end{array}\right.\,.
  \ee
With an ansatz $c(x)\equiv 1-6\frac{1}{x(x+1)}$ we find 
  \be
  x \in \left\{
  \frac{u-s}{t-r+s-u}, \frac{r-t}{t-r+s-u},
  \frac{s+u}{t+r-u-s}, \frac{t+r}{u+s-t-r}
  \right\}\,,
  \ee
i.e.\ $x\in\BQ$.
This proves our first claim that logarithmic null vectors only appear in
the framework of (quasi-)rational CFTs. The further claims follow then from 
the well known embedding structure of Verma modules for central charges with
rational $x$ (which by the way ensures $c\leq 1$ or $c\geq 25$, where at
the limiting points $x_{c\rightarrow 1}\rightarrow\infty$ and
$x_{c\rightarrow 25}\rightarrow-\frac{1}{2}$).

Obviously, null vectors in rank $r$ Jordan cells with conformal weight $h$
require the existence of $r$ different solutions $(r_i,s_i)$ such that 
$h_{r_i,s_i}(c)=h$. Up to level 5 there is only one case with $r>2$, namely
the rank 3 logarithmic null vector of the $c=c_{-1,1}=25$ theory with 
$h=h_{2,2}=h_{1,3}=h_{3,1}=-3$.

What remains is to find the numbers $r,s,t,u$ (or more generally $r_i,s_i$).
The allowed solutions must satisfy the conditions stated above:
A quadruple $(r,s,t,u)$ parametrizes a logarithmic null vector, if
with $c=c(r,s,t,u)$ one of the solutions (\ref{eq:csol}) for the central 
charge, both $h_{r,s}(c),h_{t,u}(c)\in\del H(c)$ 
where $H(c)\equiv H(x,x+1)$ is the conformal grid of the Virasoro CFT with 
central charge $c=c(x)$. This gives the conformal weights of the ``primary''
logarithmic pairs, the other possibilities are of the form $h
\in\del H(c)$ mod $\BZ_+$ and belong to ``descendant'' logarithmic pairs.
We use quotation marks because the logarithmic partner of a primary field
is not primary in the usual sense.

As an example, we consider the by now well known models with
$c=c_{p,1}$, $p>1$. Precisely all fields
in the extended conformal grid (obtained by formally considering
$c=c_{3p,3}$) except $h_{1,p}$ and $h_{1,2p}$ as well as their ``duals''
$h_{2,2p}$ and $h_{2,p}$ form tripels $(h_{1,r}=h_{1,2p-r},h_{1,2p+r})$ 
which constitute a rank 2 Jordan cell with an additional Jordan cell like
module staggerd into it (for details see \cite{Roh96}). The excluded 
fields form irreducible representations without any null vectors and are all
$\in\del^2 H(p,1)$ mod $\BZ_+$. Similar results hold for the 
$c=c_{-p,1}$, $p>1$,
models. However, all these LCFTs are only of rank 2.
The only cases of higher rank LCFTs seem to be particular $c=1$ and $c=25$
theories. Notice that such theories are necessarily {\em nonunitary\/}, i.e.\
the Shapovalov form is necessarily not positive definite. However, since
we are able to explicitly construct these theories, e.g.\ the explicit
null vectors in the Appendix, there is no doubt that these theories exist.
The reason is that the $c_{\pm p,1}$, $p>1$, theories still have additional
symmetries such that a truncation of the conformal grid to finite size still
can be constructed, while the $c=1$ and $c=25$ theories presumably are only
quasirational, their conformal grid being infinite in at least one direction.

To support our general statements, we give all non-trivial solutions up to
level 20, which are {\em not\/} $c_{p,\pm 1}$ models, in the following table. 
This means that we list only logarithmic extensions of minimal models.
The levels $r_is_i$ of the null vectors are given in decreasing order for
each central charge.  As a general fact, the null vector
with the lowest level in a given Jordan cell belongs to the irreducible
subrepresentation and is an ordinary one, all others are logarithmic.

One further comment is in order here: Minimal models may be augmented by
including the fields from the boundary of the conformal grid. However,
this alone does not suffice to get a rational LCFT. The staggered module 
structure \cite{Roh96} suggests that we also must include the fields from 
the ``next'' boundaries modulo $p,q$, i.e.\ fields with conformal
weights in $\del^k H(2p,2q)$, $k=1,2$, as well as $h_{q,s},h_{r,p}$ with 
$q<r<2q,p<s<2p$.  This is also supported by analogous results for $c_{p,1}$ 
models and their fusion rules \cite{Flo95,GaKa96}, where fusion closes if the
full (staggered) modules are considered. However, the question whether this 
really leads to a closed fusion product and therefore to rational models of
augmented minimal models is left to future work. Nonetheless, our formalism
suggests that the operator product expansion (OPE) of logarithmic fields,
as discussed in section two, closes within logarithmic fields such that there 
is a maximal rank for all Jordan cell structures. However, as concluded in
section two, the identity operator of a CFT determines the degree of its
``logarithmiticity'' because its Jordan cell structure determines which
correlation functions can be non-zero. Augmented minimal models do not
seem to have logarithmic partners of the identity field itself, but they
do have a degenerate vacuum (which thus forms a trivial diagonal Jordan cell).
This degeneracy is presumably sufficient to ensure that correlation functions
of logarithmic fields do not vanish. At least, our formalism is constructed
in such a way that it can smoothly be applied to operators $\Phi$ with
Jordan cell structure as e.g.\
${\Phi\ \lambda\choose 0\ \Phi}{\vac{h;1}\choose\vac{h;0}}$ even in the case 
$\lambda\rightarrow 0$, i.e.\ in the case that representations just have a 
multiplicity $>1$.
  $$
  \begin{array}{c|ccc}
  c & h_{r_i,s_i}\\
    \noalign{\smallskip}\hline\hline\noalign{\smallskip}
  c_{9,2}=\fr{-46}{3} & h_{2,10}=h_{2, 8}=\fr{-5}{8}     \\\noalign{\smallskip}
  c_{7,2}=\fr{-68}{7} & h_{2, 8}=h_{2, 6}=\fr{-3}{8}    
                      & h_{2, 9}=h_{2, 5}=\fr{-9}{56}  
                      & h_{2,10}=h_{2, 4}=\fr{11}{56}    \\\noalign{\smallskip}
  c_{5,2}=\fr{-22}{5} & h_{2, 6}=h_{2, 4}=\fr{-1}{8}
                      & h_{2, 7}=h_{2, 3}=\fr{7}{40}
                      & h_{3, 5}=h_{1, 5}=\fr{2}{5}      \\\noalign{\smallskip}
                      & h_{2, 8}=h_{2, 2}=\fr{27}{40}  
                      & h_{2, 9}=h_{2, 1}=\fr{11}{8}
                      & h_{2,10}=h_{4, 5}=\fr{91}{40}    \\\noalign{\smallskip}
  c_{5,3}=\fr{-3}{5}  & h_{3, 6}=h_{3, 4}=\fr{1}{12}
                      & h_{4, 5}=h_{2, 5}=\fr{7}{20}     \\\noalign{\smallskip}
  c_{3,2}=0           & h_{4, 5}=h_{2, 4}=h_{2,2}=\fr{1}{8}
                      & h_{4, 4}=h_{2, 5}=h_{2,1}=\fr{5}{8}
                      & h_{3, 6}=h_{3, 3}=h_{1,3}=\fr{1}{3}
                                                         \\\noalign{\smallskip}
                      & h_{2, 6}=h_{4, 3}=\fr{35}{24}   
                      & h_{2, 7}=h_{4, 2}=\fr{21}{8}
                      & h_{5, 3}=h_{1, 6}=\fr{10}{3}     \\\noalign{\smallskip}
                      & h_{2, 8}=h_{4, 1}=\fr{33}{8}
                      & h_{2, 9}=h_{6, 3}=\fr{143}{24}   
                      & h_{2,10}=h_{6, 2}=\fr{65}{8}     \\\noalign{\smallskip}
  c_{4,3}=\fr{1}{2}   & h_{3, 5}=h_{3, 3}=\fr{1}{6}
                      & h_{4, 4}=h_{2, 4}=\fr{5}{16}
                      & h_{3, 6}=h_{3, 2}=\fr{35}{48}    \\\noalign{\smallskip}
                      & h_{5, 4}=h_{1, 4}=\fr{21}{16}    \\\noalign{\smallskip}
  c_{4,-3}=\fr{51}{2} & h_{5, 4}=h_{2, 8}=\fr{-325}{16}
                      & h_{3, 6}=h_{6, 2}=\fr{-851}{48}
                      & h_{4, 4}=h_{1, 8}=\fr{-245}{16}  \\\noalign{\smallskip}
                      & h_{3, 5}=h_{6, 1}=\fr{-85}{6}    \\\noalign{\smallskip}
  c_{3,-2}=26         & h_{2,10}=h_{8, 1}=\fr{-217}{8}
                      & h_{2, 9}=h_{6, 3}=\fr{-551}{24}
                      & h_{4, 5}=h_{2, 8}=h_{6,2}=\fr{-153}{8}
                                                         \\\noalign{\smallskip}
                      & h_{3, 6}=h_{5, 3}=h_{1,9}=\fr{-52}{3}
                      & h_{4, 4}=h_{2, 7}=h_{6,1}=\fr{-125}{8}
                      & h_{2, 6}=h_{4, 3}=\fr{-299}{24}  \\\noalign{\smallskip}
                      & h_{2, 5}=h_{4, 2}=\fr{-77}{8}
                      & h_{3, 3}=h_{1, 6}=\fr{-25}{3}   
                      & h_{2, 4}=h_{4, 1}=\fr{-57}{8}    \\\noalign{\smallskip}
  c_{5,-3}=\fr{133}{5}& h_{4, 5}=h_{1,10}=\fr{-387}{20}
                      & h_{3, 6}=h_{6, 1}=\fr{-205}{12}  \\\noalign{\smallskip}
  c_{5,-2}=\fr{152}{5}& h_{2,10}=h_{4, 5}=\fr{-851}{40}
                      & h_{2, 9}=h_{4, 4}=\fr{-147}{8}
                      & h_{2, 8}=h_{4, 3}=\fr{-627}{40}  \\\noalign{\smallskip}
                      & h_{3, 5}=h_{1,10}=\fr{-72}{5}    
                      & h_{2, 7}=h_{4, 2}=\fr{-527}{40}
                      & h_{2, 6}=h_{4, 1}=\fr{-87}{8}    \\\noalign{\smallskip}
  c_{7,-2}=\fr{250}{7}& h_{2,10}=h_{4, 3}=\fr{-1075}{56}
                      & h_{2, 9}=h_{4, 2}=\fr{-943}{56}
                      & h_{2, 8}=h_{4, 1}=\fr{-117}{8}   \\\noalign{\smallskip}
  c_{9,-2}=\fr{124}{3}& h_{2,10}=h_{4, 1}=\fr{-147}{8}
  \end{array}
  $$

\subsubsection*{{\em The\/ $(h,c)$ Plane: Null Vectors with Level\/ 
$rs\leq 400$.}}

\nn It might be illuminating, and the author is fond of plots anyway,
to plot the sets $\del^k H(p,q)$, $k=0,1,2$, for a variety of CFTs. 
The product $pq$ is roughly a measure for the size of the CFT since the
size of the conformal grid and thus the field content is determined by it.
Thus, it seems reasonable to plot all sets with $pq\leq n$ where we have 
chosen $n=400$. 

To make the structure of the $(h,c)$ plane better visible,
we transformed the variables via 
  \be\label{eq:coord}
  x\mapsto {\rm sign}(x)\log(|x|+1)\ {\rm for}\ x=h,c\,,
  \ee
which amounts in a double logarithmic scaling of the axes both, in positive
as well as in negative direction. The conformal weights are plotted in 
horizontal direction, the central charges along the vertical direction.
The following plots show only the part
of the $(h,c)$ plane which belongs to $c\leq 1$ CFTs, i.e.\ minimal models
and $c_{p,1}$ LCFTs, $p>0$. The other ``half'' with $c\geq 25$ shares 
analogous features. Due to the map (\ref{eq:coord}) the vertical range of 
roughly $[-5.5,1.0]$ coresponds to $-240\leq c\leq 1$, whereas the horizontal
range $[-5.5,5.0]$ does roughly correspond to $-240\leq h\leq 148$. Actually,
the lables $\pm\{0,1,2,3,4,5\}$ correspond to values $h,c=
\pm\{0,1.718,6.389,19.086,53.598,147.413\}$, given here for better orientation. 

We did not plot $(\del^2 H(p,q),c_{p,q})$ since these points just lie
at the left border of the point-set in figure 1, they belong to
the highest weight representations with type $III_{\pm}^{\circ\circ}$
embedding structure of Verma modules.
If one would put both plots above each other, 
one might infer from them that the set of logarithmic representations
precisely lies on the ``forbidden'' curves of the pointset of ordinary
highest weight representations. This illustrates the fact that logarithmic
representations appear, if the conformal weights of two highest weight
representations become identical. 

As discussed in our earlier work \cite{Flo95}, this
situation arises in the limit of series of minimal models
$c_{p_1,q_1}, c_{p_2,q_2}, c_{p_3,q_3}, \ldots$ with 
$\lim_{i\rightarrow\infty}p_iq_i=\infty$. Usually, the field content of
these theories increases with $i$, but it might happen that in the limit
$p_i$ and $q_i$ become almost coprime. More precisely, a sequence such as
for example $\{c_{\alpha p,(\alpha+1)q}\}_{\alpha\in\BZ_+}$ converges
to a limiting theory with central charge 
$\lim_{\alpha\rightarrow\infty}c_{\alpha p,(\alpha+1)q} = c_{p,q}$. Therefore,
we expect a rather small field content at the limit point since the conformal 
weights of the $c_{\alpha p,(\alpha+1)q}$ theories also approach the ones of 
the $c_{p,q}$ model (modulo $\BZ$). A more detailed analysis (second reference
in \cite{Flo95}) reveals that indeed conformal weights approach each other 
giving rise for Jordan cells. Hence, the theory at the limit point, while 
having central chrage $c_{p,q}$ actually is a LCFT.
The plots presented here clearly visualize this topology of the space of CFTs
in the $(h,c)$ plane of their spectra.

To summarize, our results strongly suggest that augmented minimal models
form {\em rational\/} logarithmic conformal field theories in the same
sense as the $c_{p,1}$ models do. The only difference between the former and 
the latter is that for the $c_{p,1}$ models $H(p,1) = \emptyset$. We know since
BPZ \cite{BPZ83} that under fusion $H(p,q)\times H(p,q) \rightarrow H(p,q)$,
and since \cite{GaKa96,Flo95} that under fusion $H'(p,q)\times H'(p,q)
\rightarrow H'(p,q)$ with $H'(p,q)=\del H(p,q)\cup\del^2 H(p,q)$, if we deal
with the full indecomposable representations. Therefore, the only difficulty 
can come from mixed fusion products of type $H(p,q)\times H'(p,q)$ which
traditionally (without logarithmic operators) are zero due to decoupling.
The formalism presented in section 2, however, yields non-zero fusion products
by paying the price that representations from $H(p,q)$ appear with non-trivial
multiplicities (because of the fact that the corresponding OPEs yield 
fields on the right hand side with $h\in H(p,q)$ mod $\BZ$, which have a 
non-trivial dependence on the formal $\theta$ variables.
As mentioned before, a more detailed analysis of augmented 
minimal models is left for future work.

  \vfill\pn
{\bf Acknowledgment:} ~~I would like to thank Ralph Blumenhagen, Victor 
Gurarie and Ian Kogan for discussions and comments. I am also very grateful 
to Ralf Krafft. This work has been supported by the Deutsche 
Forschungsgemeinschaft.
  \vfill

\fig{{\em ~~All the spectra $(H(p,q),c_{p,q})$ for all $p,q>0$ coprime 
such that $pq\leq 400$, which constitutes the set of all irreducible highest 
weight representations of minimal models. This means that for each central
charge $c$ we plotted all conformal weights $h$ from within the truncated 
conformal grid $\{h_{r,s}:0<r<q,0<s<p\}$. See text for details about the
logarithmic scaling (equation \ref{eq:coord}).}}{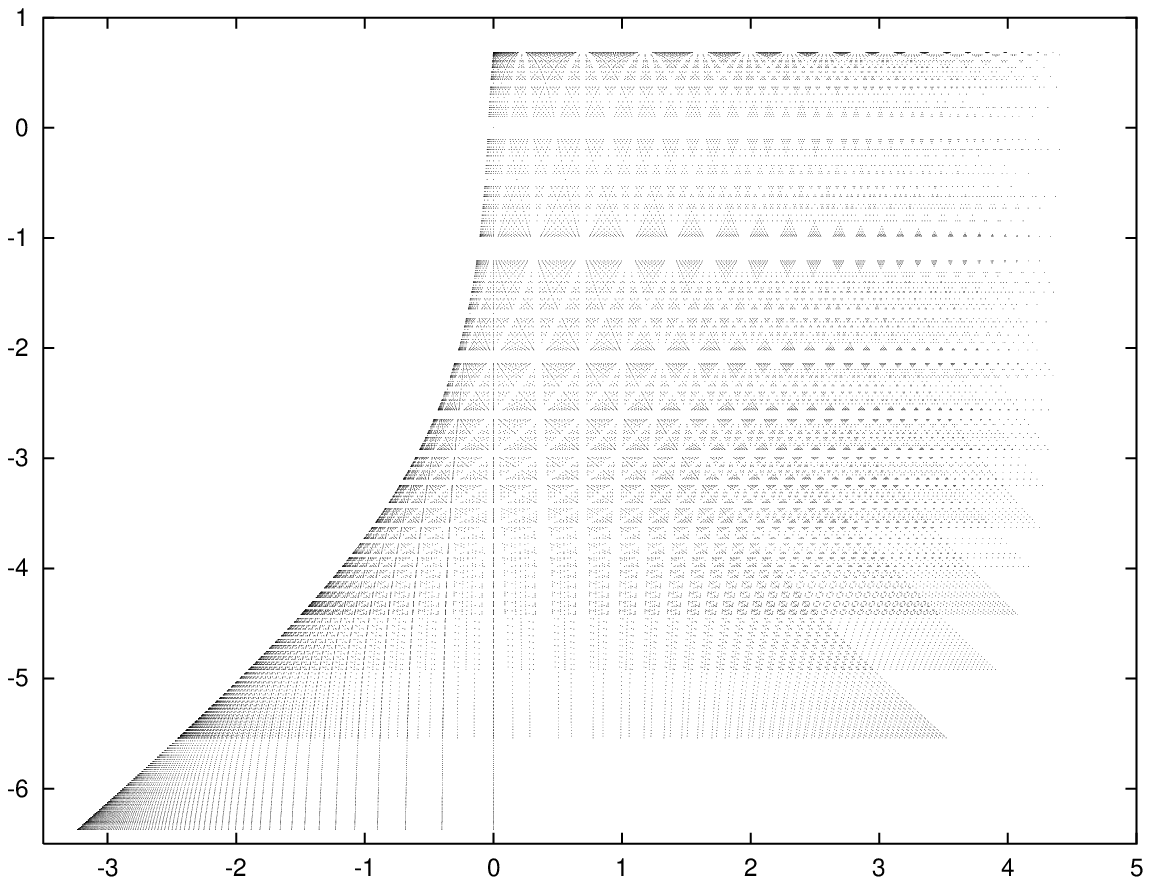}{15truecm}

\fig{{\em ~~All the spectra $(\del H(p,q),c_{p,q})$ for all $p,q>0$ coprime 
such that $pq\leq 400$, which constitutes the set of all Jordan cell 
representations, i.e.\ all conformal weights where fields with logarithmic 
partners exist. This means that for each central charge $c$ we plotted all
conformal weights $h$ from the boundary of the truncated conformal grid
$\{h_{q,s}:0<s\leq p\}\cup\{h_{r,p}:0<r\leq q\}$.
See text for details about the logarithmic 
scaling (equation \ref{eq:coord}).}}{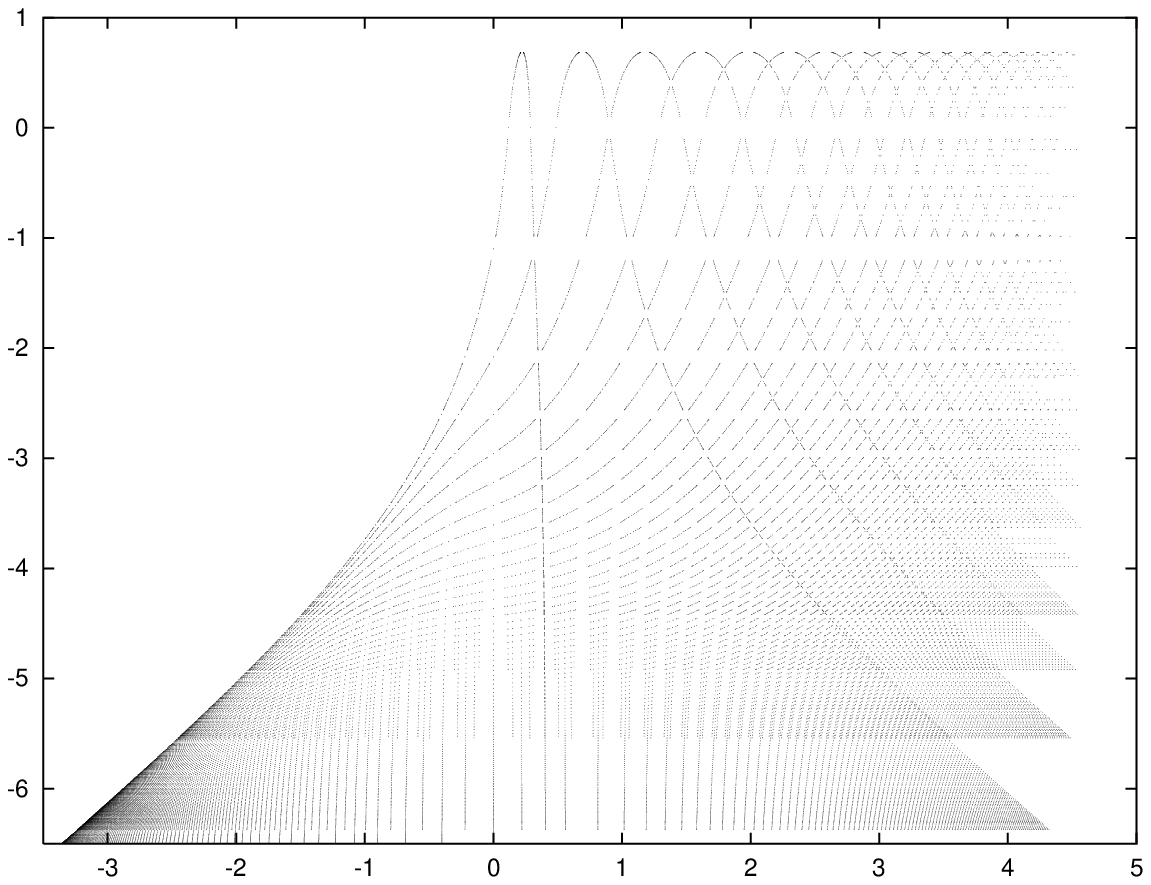}{15truecm}

  \newpage

%
%%< APPENDIX A >%%%%%%%%%%%%%%%%%%%%%%%%%%%%%%%%%%%%%%%%%%%%%%%%%%%%%%%%%%%
%
  \appendix
  \renewcommand{\theequation}{\mbox{\Alph{section}.\arabic{equation}}}
  \mysection{Appendix}

\nn We present explicit results for all possible Virasoro logarithmic null 
vectors
up to level 5. The null vector at level 1 is trivial, $\vac{\chi^{(1)}_{h=0,c}}
= L_{-1}\vac{0}$. There can be no logarithmic null vector at this level.
For null vectors of level $n>1$ we make the general ansatz
  \be
  \vac{\chi^{(n)}_{h,c}} = \sum_j\sum_{|\vec{n}|=n}b^{\vec{n}}_j(h,c)
  L_{-\vec{n}}\Vac{h;\del_{\theta}^ja(\theta)}
  \ee
and define matrix elements
  \bea
  N_{k,l}^{(n)} &=& \left.\frac{\del^k}{\del\theta^k}\left(
  \sum_j\sum_{|\vec{n}|=n}b^{\vec{n}}_j(h,c)
  \Avac{h\vphantom{\del_{\theta}^k}}L_{\vec{n}'_l}L_{-\vec{n}}
  \Vac{h;\del_{\theta}^ja(\theta)}
  \right)\right|_{\theta=0}
  \nonumber\\
          &=& \sum_{j=0}^k\sum_{|\vec{n}|=n}b^{\vec{n}}_j(h,c)
  \frac{1}{j!}\frac{\del^j}{\del h^j}
  \Avac{h}L_{\vec{n}'_l}L_{-\vec{n}}\Vac{h}\,,
  \eea
where $\vec{n}'_l$ is some enumeration of the $p(n)$ different partitions of
$n$. Since the maximal possible rank of a Jordan cell representation which
may contain a logarithmic null vector is $r\leq p(n)$, we consider $N^{(n)}$ 
to be
a $p(n)\times p(n)$ square matrix. Our particular ansatz is conveniently chosen
to simplify the action of the Virasoro modes on Jordan cells. Notice, that
the derivatives with respect to the conformal weight $h$ do not act on the
coefficients $b^{\vec{n}}_j(h,c)$. Of course, we assume that $a(\theta)$ has 
maximal degree in $\theta$, i.e.\ ${\rm deg}(a(\theta)) = r-1$.

\subsubsection*{{\em The Logarithmic Null Vectors at Level 2.}}

\nn As an example, at level 2 we have $p(2)=2$ and the matrix $N^{(2)}$ reads
  \be
  N^{(2)} = \left[\begin{array}{cc} 
  b^{\{1,1\}}_0\left(8h^2+4h\right)+6b^{\{2\}}_0h &
  b^{\{1,1\}}_0\left(16h+4\right)+6b^{\{2\}}_0
  +b^{\{1,1\}}_1\left(8h^2+4h\right)+6b^{\{2\}}_1h\\\noalign{\medskip}
  6b^{\{1,1\}}_0h+b^{\{2\}}_1\left(4h+\fr{1}{2}c\right) &
  6b^{\{1,1\}}_0+4b^{\{2\}}_1+6b^{\{1,1\}}_1h
  +b^{\{2\}}_1\left(4h+\fr{1}{2}c\right)
  \end{array}\right]\,.
  \ee
A null vector is logarithmic of rank $k\geq 0$ if the first $k+1$ columns
of $N^{(n)}$ are zero, where $k=0$ means an ordinary null vector. As described 
in
the text, one first solves for ordinary null vectors (such that the first
column vanishes up to one entry). This determines the $b^{\vec{n}}_0(h,c)$.
Then one puts $b^{\vec{n}}_k(h,c) = \frac{1}{k!}\del_h^{k}
b^{\vec{n}}_0(h,c)$. Without loss of generality we may then assume that all
entries except the last row are zero. In our example, this procedure 
results in
  \be
  N^{(2)} = \left[\begin{array}{cc}
  0 & 0 \\\noalign{\medskip}
  10h^2-16h^3-2h^2c-hc & 20h-48h^2-4hc-c
  \end{array}\right]\,,
  \ee
where $b^{\{1,1\}}_k = \frac{1}{k!}\del_h^{k}(3h)$ and
$b^{\{2\}}_k = \frac{1}{k!}\del_h^{k}(-2h(2h+1))$ upto an overall 
normalization. The last step is trying to find simultanous solutions
for the last row, i.e.\ common zeros of polynomials $\in\BC[h,c]$. In our
example, $N^{(2)}_{2,1}=0$ yields $c = 2h(5-8h)/(2h+1)$. Then, the last
condition becomes $N^{(2)}_{2,2} = -2h(16h^2+16h-5)/(2h+1) = 0$ which can 
be satisfied for $h\in\{0,\frac{-5}{4},\frac{1}{4}\}$. From this we finally
obtain the explicit logarithmic null vectors at level 2:
  $$
  \begin{array}{c|c}
  (h,c) & \vac{\chi^{(2)}_{h,c}}
    \\\noalign{\smallskip}\hline\hline\noalign{\smallskip}
  (0,0) & 
    (3L_{-1}^2-2L_{-2})\Vac{0;a(\theta)}
    \\\noalign{\smallskip}
  (\frac{1}{4},1) &
    (3L_{-1}^2-3L_{-2})\Vac{\frac{1}{4};a(\theta)} 
    -4L_{-2}\Vac{\frac{1}{4};\del_{\theta}a(\theta)}
    \\\noalign{\smallskip}
  (\frac{-5}{4},25) & 
    (3L_{-1}^2+3L_{-2})\Vac{\frac{-5}{4};a(\theta)} 
    -4L_{-2}\Vac{\frac{-5}{4};\del_{\theta}a(\theta)}
  \end{array}
  $$
Note, that according to our formalism, $h=0,c=0$ does not turn out to be
a logarithmic null vector at level 2. Here and in the following the highest 
order derivative $\del_{\theta}^k a(\theta)$ indicates the maximal rank
of a logarithmic null vector to be $k$ (and hence the maximal rank of the 
corresponding Jordan cell representation to be $r=k+1$). It is implicitly
understood that $a(\theta)$ is then chosen such that the highest order
derivative yields a non-vanishing constant.

All null vectors are normalized such that all coefficients are integers.
Clearly, they are not unique since with $\Vac{\chi(\theta)} =
\sum_k \Vac{\chi^{}_k;\del_{\theta}^k a(\theta)}$ every vector
  \be
  \Vac{\chi'(\theta)} = \sum_k 
  \Vac{\chi^{}_k;\sum_{l\geq 0}\lambda_{k,l}\del_{\theta}^{k+l}a(\theta)}
  \ee
is also a null vector.

\subsubsection*{{\em The Logarithmic Null Vectors at Level 3.}}

\nn Following the same procedure for level 3 null vectors, we first obtain two
branches of solutions $c=c_i(h)$, namely 
  \be
  c_1(h) = -2h\frac{8h-5}{1+2h}\,,\ \ \ \
  c_2(h) = -\frac{3h^2-7h+2}{h+1}\,.
  \ee
Clearly, $c_1(h)$ is the same solution as obtained at level 2, meaning the
trivial fact that each level 2 null vector is also a null vector at level 3.
On the other hand, these solutions need not all be redundant, because it might
happen that a null vector at level $n$, which turned out to be just an
ordinary one, becomes a logarithmic null vector at a higher level $n'>n$.
The results for level 3 are:
  $$
  \begin{array}{c|c}
  (h,c) & \vac{\chi^{(3)}_{h,c}}\ph{\rm(trivial\ solutions)}
    \\\noalign{\smallskip}\hline\hline\noalign{\smallskip}
  (\frac{-1}{5},\frac{-22}{5}) &
    (175L_{-1}^3+350L_{-2}L_{-1}-154L_{-3})\Vac{\frac{-1}{5};a(\theta)}
    \\\noalign{\smallskip}
  (0,0) &
    (-3L_{-1}^3+2L_{-2}L_{-1}+2L_{-3})\Vac{0;a(\theta)}
    \\\noalign{\smallskip}
  (\frac{1}{2},\frac{1}{2}) &
    (14L_{-1}^3-28L_{-2}L_{-1}-7L_{-3})\Vac{\frac{1}{2};a(\theta)}
    \\\noalign{\smallskip}
  \end{array}
  $$
  $$
  \begin{array}{c|c}
  (h,c) & \vac{\chi^{(3)}_{h,c}}\ph{\rm(logarithmic\ null\ vectors)}
    \\\noalign{\smallskip}\hline\hline\noalign{\smallskip}
  (0,-2) &
    (-L_{-1}^3+2L_{-2}L_{-1})\Vac{0;a(\theta)}
    -L_{-3}\Vac{0;\del_{\theta}a(\theta)}
    \\\noalign{\smallskip}
  (\frac{1}{4},1) &
    (-9L_{-1}^3+9L_{-2}L_{-1}+9L_{-3})\Vac{\frac{1}{4};a(\theta)}
    \\\noalign{\smallskip} & 
    +\,(28L_{-1}^3-16L_{-2}L_{-1}-16L_{-3})
    \Vac{\frac{1}{4};\del_{\theta}a(\theta)}
    \\\noalign{\smallskip}
  (1,1) &
    (3L_{-1}^3-12L_{-2}L_{-1}+6L_{-3})\Vac{1;a(\theta)}
    \\\noalign{\smallskip} & 
    +\,(7L_{-1}^3-34L_{-2}L_{-1}+23L_{-3})\Vac{1;\del_{\theta}a(\theta)}
    \\\noalign{\smallskip} 
  (-3,25) &
    (-49L_{-1}^3-196L_{-2}L_{-1}-294L_{-3})\Vac{-3;a(\theta)}
    \\\noalign{\smallskip} & 
    +\,(7L_{-1}^3+126L_{-2}L_{-1}+287L_{-3})\Vac{-3;\del_{\theta}a(\theta)}
    \\\noalign{\smallskip} 
  (\frac{-5}{4},25) & 
    (-147L_{-1}^3-147L_{-2}L_{-1}-147L_{-3})\Vac{\frac{-5}{4};a(\theta)}
    \\\noalign{\smallskip} & 
    +\,(28L_{-1}^3+224L_{-2}L_{-1}+224L_{-3})
    \Vac{\frac{-5}{4};\del_{\theta}a(\theta)}
    \\\noalign{\smallskip}
  (-2,28) &
    (-45L_{-1}^3-90L_{-2}L_{-1}-90L_{-3})\Vac{-2;a(\theta)}
    \\\noalign{\smallskip} & 
    +\,(7L_{-1}^3+86L_{-2}L_{-1}+104L_{-3})\Vac{-2;\del_{\theta}a(\theta)}
  \end{array}
  $$
{}From this we can infer that the first three solutions are only ordinary
null vectors, and that all remaining solutions correspond to rank 2 Jordan
cell highest weight representations. We also find the solutions already
obtained at level 2, as is to be expected. Some care has to be taken with the
solution $h=0,c=-2$. The $c=c_{2,1}=-2$ model is well known and
has a rank 2 Jordan cell structure for the $h=0$ representation. So, we would
expect a logarithmic null vector of level 3, since the logarithmic partner
of the primary field $\Psi_{(h=0;0)}(z)\equiv\Phi_{1,1}(z)=\Bbb{I}$ is the 
field $\Psi_{(h=0;1)}(z)\equiv\Phi_{1,3}(z)$. However, our standard
ansatz does not reveal this null vector. The secret of this special solution
is that the irreducible sub-representation at $h=0$ is the {\em vacuum\/} 
representation, which always contains a level 1 null vector. Therefore, we
have the additional freedom to add a ``zero'' via
  \be
  \Vac{\chi^{(3)}_{h=0,c}} \mapsto
  \Vac{\chi^{(3)}_{h=0,c}} + \left(\alpha L_{-1}^2+\beta L_{-2}\right)L_{-1}
  \Vac{0;\del_{\theta}a(\theta)}\,.
  \ee
Moreover, such special solutions with $h\!=\!0$ do not satisfy our proposed
general relation that $\left.b^{\vec{n}}_1(h,c)
\right|_{h=0,c=c_i(0)}=\left.\del_h b^{\vec{n}}_0(h,c)\right|_{h=0,c=c_i(0)}$
with $c=c_i(h)$ the appropriate general solution for the central charge for 
generic $h$ (note that the level 1 null vector in the vacuum representation
is independent of $c$).
The reason for this is that for $h\neq 0$ we can use $h$ as a homogeneous
coordinate in the projective space of states (due to the freedom of
normalization). For $h=0$ the only coordinate left is $c$ and hence
we must calculate
  \be
  \left.b^{\vec{n}}_k(h,c)\right|_{h=0,c=c_i(0)} = 
  \left.\frac{1}{k!}\del_h^k b^{\vec{n}}_0(h,c_i(h))\right|_{h=0}
  \ee
instead, where the central charge has been replaced in advance by the
appropriate solution $c=c_i(h)$ for generic $h$. 

It follows, that under certain circumstances $h\!=\!0$ representations may be 
extended to Jordan cell structure of rank at least 2. This is expected due to
the fact that all $c_{p,1}$ rational LCFTs have a rank 2 Jordan cell 
representation containing the vacuum (identity) representation. However,
these representations have a more complicated structure, because they form
together with a representation on $h\!=\!1$ a staggered indecomposable module
\cite{Roh96}.

The recipe to find such null vectors with $h\!=\!0$ is simple. One observes 
that
$L_0\vac{0;k}=(1-\delta_{k,0})\vac{0;k-1}$. It is then easy to see that the
only non-trivially vanishing matrix elements at level $n$ are of the form
$\avac{0;a'(\theta')}L_{\vec{n}'}L_{-\vec{n}}\vac{0;a(\theta)}$ where both
partitions $\vec{n}'=\{n'_1,n'_2,\ldots\}$, $\vec{n}=\{n_1,n_2,\ldots\}$ do
not contain 1. Restricting ourselves to the case of rank 2 Jordan
cells, and putting the coefficients $b^{\vec{n}}_0(0,c)$ to zero where
$\vec{n}$ does not contain 1, we are left solely with conditions
for $b^{\vec{n}}_1(0,c)$, $1\not\in\vec{n}$, and $c$, which depend on the
$b^{\vec{n}}_0(0,c)$, $1\in\vec{n}$. For $n=2$ we would have to find a
solution for $b^{\{2\}}_1$ and $c$ which both depend only on one variable,
$b^{\{1,1\}}_0$, so it is not surprising that we didn't find any. For $n=3$
however, we again have two constraints ($b^{\{3\}}_1$, $c$), but now
also two variables ($b^{\{1,1,1\}}_0$, $b^{\{2,1\}}_0$). For $n=4$ we have
three constraints and three variables etc., in general there are
$p_2(n) + 1$ constraints and $p(n)-p_2(n)$ variables, where $p_2(n)$ is
the number of partitions of $n$ not containing 1.

\subsubsection*{{\em The Logarithmic Null Vectors at Level 4.}}

\nn The next results are given without further comments and with trivial (non
logarithmic) solutions omitted. Level $n=4$ is the smallest level where
we find a logarithmic null vector of rank 3, namely with $h=-3,c=25$.

{\vskip-0.75ex\footnotesize
  $$
  \begin{array}{c|c}
  (h,c) & \vac{\chi^{(4)}_{h,c}}
    \\\noalign{\smallskip}\hline\hline\noalign{\smallskip}
  (-\frac{1}{4},-7) & 
    (315L_{-1}^4+315L_{-2}^2-210L_{-3}L_{-1}-210L_{-4}-1050L_{-2}L_{-1}^2)
    \Vac{\frac{-1}{4};a(\theta)} 
    \\\noalign{\smallskip} &  
    +\,(-878L_{-3}L_{-1}+2577L_{-1}^4-11830L_{-2}L_{-1}^2+3657L_{-2}^2
    -1718L_{-4})\Vac{\frac{-1}{4};\del_{\theta}a(\theta)}
    \\\noalign{\smallskip}
  (0,-2) &
    (L_{-1}^4 - 2L_{-2}L_{-1}^2 - 2L_{-3}L_{-1})\Vac{0;a(\theta)}
    +2L_{-4}\Vac{0;\del_{\theta}a(\theta)}
    \\\noalign{\smallskip}
  (\frac{3}{8},-2) &
    (1260L_{-1}^4+2835L_{-2}^2+1260L_{-3}L_{-1}-1890L_{-4}-6300L_{-2}L_{-1}^2)
    \Vac{\frac{3}{8};a(\theta)}
    \\\noalign{\smallskip} & 
    +\,(3832L_{-3}L_{-1}+2152L_{-1}^4-14120L_{-2}L_{-1}^2+9882L_{-2}^2
    -7008L_{-4})\Vac{\frac{3}{8};\del_{\theta}a(\theta)}
    \\\noalign{\smallskip}
  (0,1) &
    (-3L_{-1}^4 + 12L_{-2}L_{-1}^2 - 6L_{-3}L_{-1})\Vac{0;a(\theta)}
    +(-16L_{-2}^2+12L_{-4})\Vac{0;\del_{\theta}a(\theta)}
    \\\noalign{\smallskip}
  (1,1) &
    (-60L_{-1}^4+240L_{-2}L_{-1}^2+120L_{-3}L_{-1}-240L_{-4})\Vac{1;a(\theta)}
    \\\noalign{\smallskip} & 
    +\,(-89L_{-1}^4+476L_{-2}L_{-1}^2+118L_{-3}L_{-1}-716L_{-4})
    \Vac{1;\del_{\theta}a(\theta)}
    \\\noalign{\smallskip}
  (\frac{9}{4},1) &
    (45L_{-1}^4+405L_{-2}^2+630L_{-3}L_{-1}-810L_{-4}-450L_{-2}L_{-1}^2)
    \Vac{\frac{9}{4};a(\theta)}
    \\\noalign{\smallskip} & 
    +\,(1996L_{-3}L_{-1}+110L_{-1}^4-1220L_{-2}L_{-1}^2+1206L_{-2}^2
    -2772L_{-4})\Vac{\frac{9}{4};\del_{\theta}a(\theta)}
    \\\noalign{\smallskip}
  (-\frac{21}{4},25) &  
    (-990L_{-1}^4-8910L_{-2}^2-33660L_{-3}L_{-1}-65340L_{-4}
    -9900L_{-2}L_{-1}^2)\Vac{\frac{-21}{4};a(\theta)}   
    \\\noalign{\smallskip} &     
    \,+(45946L_{-3}L_{-1}+901L_{-1}^4+11650L_{-2}L_{-1}^2+12861L_{-2}^2
    +102234L_{-4})\Vac{\frac{-21}{4};\del_{\theta}a(\theta)}
    \\\noalign{\smallskip}
  (-3,25) &
    (63504L_{-1}^4+254016L_{-2}L_{-1}^2+635040L_{-3}L_{-1}+762048L_{-4})
    \Vac{-3;a(\theta)}
    \\\noalign{\smallskip} & 
    +\,(59283L_{-1}^4+110124L_{-2}L_{-1}^2+148302L_{-3}L_{-1}+76356L_{-4})
    \Vac{-3;\del_{\theta}a(\theta)}
    \\\noalign{\smallskip} &
    +\,(-15104L_{-1}^4-186920L_{-2}L_{-1}^2-63504L_{-2}^2-450920L_{-3}L_{-1}
    -575628L_{-4})\Vac{-3;\del_{\theta}^2a(\theta)}
    \\\noalign{\smallskip}
  (-\frac{27}{8},28) &
    (77220L_{-1}^4+173745L_{-2}^2+849420L_{-3}L_{-1}+1042470L_{-4}
    +386100L_{-2}L_{-1}^2)\Vac{\fr{-27}{8};a(\theta)}
    \\\noalign{\smallskip} &
    +\,(269896L_{-3}L_{-1}+71336L_{-1}^4+150760L_{-2}L_{-1}^2-148374L_{-2}^2
    +113616L_{-4})\Vac{\fr{-27}{8};\del_{\theta}a(\theta)}
    \\\noalign{\smallskip}
  (-2,28) &
    (13860L_{-2}L_{-1}^2+27720L_{-3}L_{-1}+27720L_{-4}+6930L_{-1}^4)
    \Vac{-2;a(\theta))}
    \\\noalign{\smallskip} & 
    +\,(1577L_{-1}^4-9716L_{-2}L_{-1}^2-3564L_{-2}^2-18640L_{-3}L_{-1}
    -21412L_{-4})\Vac{-2;\del_{\theta}a(\theta)}
    \\\noalign{\smallskip}
  (-\frac{11}{4},33) &
    (208845L_{-1}^4+696150L_{-2}L_{-1}^2+208845L_{-2}^2+1253070L_{-3}L_{-1}
    +1253070L_{-4})\Vac{\fr{-11}{4};a(\theta)}
    \\\noalign{\smallskip} &
    +\,(58354L_{-1}^4-244540L_{-2}L_{-1}^2-304086L_{-2}^2-525036L_{-3}L_{-1}
    -684156L_{-4})\Vac{\fr{-11}{4};\del_{\theta}a(\theta)}
  \end{array}
  $$}

\subsubsection*{{\em The Logarithmic Null Vectors at Level 5.}}

\nn Again, we only list the non-trivial results. Searching for logarithmic
null vectors produces a high amount of trivial solutions (mostly from
minimal models). Indeed, although minimal models contain pairs of operators 
$\Phi_{r,s}(z)$, $\Phi_{t,u}(z)$ with $h_{r,s}(c)=h_{t,u}(c)$, these operators
are not different ones, but are identified with each other. Therefore, 
the existence of logarithmic operators is bound to the non-existence of a
BRST operator such that these pairs of operators get identified in the
BRST invariant representation modules. It has been shown for the $c_{p,1}$ 
models that one of the BRST charges becomes a local operator and thus an
element of the field content of the theory itself, spoiling the usual
field identifications in minimal models. 

{\vskip-0.75ex\footnotesize
  $$ 
  \begin{array}{c|l}
  (h,c) & \vac{\chi^{(5)}_{h,c}}
    \\\noalign{\smallskip}\hline\hline\noalign{\smallskip}
  (-\frac{1}{2},-\frac{25}{2}) &
    (168L_{-1}^5-840L_{-2}L_{-1}^3-378L_{-3}L_{-1}^2+(-504L_{-4}
    +672L_{-2}^2)L_{-1}+168L_{-3}L_{-2}
    \\\noalign{\smallskip} & 
      \ph-84L_{-5})\Vac{\fr{-1}{2};a(\theta)}
    \\\noalign{\smallskip} & 
    +\,(-260L_{-1}^5-6316L_{-2}L_{-1}^3+1425L_{-3}L_{-1}^2+(6576L_{-2}^2
    +864L_{-4})L_{-1}+2076L_{-5}
    \\\noalign{\smallskip} & 
      \ph+1308L_{-3}L_{-2})\Vac{\fr{-1}{2};\del_{\theta}a(\theta)}
    \\\noalign{\smallskip} 
  (-\frac{1}{4},-7) &
    (-4095L_{-1}^5+13650L_{-2}L_{-1}^3+16380L_{-3}L_{-1}^2+(8190L_{-4}
    -4095L_{-2}^2)L_{-1}-8190L_{-3}L_{-2}
    \\\noalign{\smallskip} & 
      \ph+4095L_{-5})\Vac{\fr{-1}{4};a(\theta)}
    \\\noalign{\smallskip} & 
    +\,(-13641L_{-1}^5+86150L_{-2}L_{-1}^3+78564L_{-3}L_{-1}^2+(-27201L_{-2}^2
    +5442L_{-4})L_{-1}+81L_{-5}
    \\\noalign{\smallskip} & 
      \ph-52482L_{-3}L_{-2})\Vac{\fr{-1}{4};\del_{\theta}a(\theta)}
    \\\noalign{\smallskip} 
  (0,-7) & (9L_{-1}^5-60L_{-2}L_{-1}^3+64L_{-2}^2L_{-1}-6L_{-3}L_{-1}^2
    -36L_{-4}L_{-1})\Vac{0;a(\theta)}
    \\\noalign{\smallskip} & 
    +\,(-32L_{-3}L_{-2}+12L_{-5})\Vac{0;\del_{\theta}a(\theta)}
    \\\noalign{\smallskip} 
  (0,-2) & (-L_{-1}^5+9L_{-2}L_{-1}^3-14L_{-2}^2L_{-1}+4L_{-3}L_{-1}^2
    +4L_{-4}L_{-1})\Vac{0;a(\theta)}
    \\\noalign{\smallskip} & 
    +\,(7L_{-3}L_{-2}+L_{-5})\Vac{0;\del_{\theta}a(\theta)}
    \\\noalign{\smallskip}
  (\frac{3}{8},-2) &
    (-13860L_{-1}^5+69300L_{-2}L_{-1}^3+55440L_{-3}L_{-1}^2+(-6930L_{-4}
    -31185L_{-2}^2)L_{-1}-62370L_{-3}L_{-2}
    \\\noalign{\smallskip} & 
      \ph+31185L_{-5})\Vac{\fr{3}{8};a(\theta)}
    \\\noalign{\smallskip} & 
    +\,(32L_{-1}^5+36800L_{-2}L_{-1}^3+18352L_{-3}L_{-1}^2+(-55368L_{-2}^2
    +4636L_{-4})L_{-1}+69228L_{-5}
    \\\noalign{\smallskip} & 
      \ph-110736L_{-3}L_{-2})\Vac{\fr{3}{8};\del_{\theta}a(\theta)}
    \\\noalign{\smallskip} 
  (1,-2) &
    (90L_{-1}^5-900L_{-2}L_{-1}^3+540L_{-3}L_{-1}^2+(-1080L_{-4}
    +1440L_{-2}^2)L_{-1}-720L_{-3}L_{-2}
    \\\noalign{\smallskip} & 
      \ph+360L_{-5})\Vac{1;a(\theta)}
    \\\noalign{\smallskip} & 
    +\,(433L_{-1}^5-3910L_{-2}L_{-1}^3+1878L_{-3}L_{-1}^2+(6568L_{-2}^2
    -6276L_{-4})L_{-1}+1012L_{-5}
    \\\noalign{\smallskip} & 
      \ph-3464L_{-3}L_{-2})\Vac{1;\del_{\theta}a(\theta)}
    \\\noalign{\smallskip}
  (0,1) & (-3L_{-1}^5+12L_{-2}L_{-1}^3+6L_{-3}L_{-1}^2-12L_{-4}L_{-1})
    \Vac{0;a(\theta)}\hfill
    +(-32L_{-3}L_{-2}+20L_{-5})\Vac{0;\del_{\theta}a(\theta)}
    \\\noalign{\smallskip}
  (\frac{1}{4},1) &
    (-7785L_{-1}^5+9450L_{-2}L_{-1}^3+48060L_{-3}L_{-1}^2+(46710L_{-4}
    -1665L_{-2}^2)L_{-1}-26370L_{-3}L_{-2}
    \\\noalign{\smallskip} & 
      \ph+45045L_{-5})\Vac{\fr{1}{4};a(\theta)}
    \\\noalign{\smallskip} & 
    +\,(18058L_{-1}^5-49260L_{-2}L_{-1}^3-24168L_{-3}L_{-1}^2+(39362L_{-2}^2
    -46068L_{-4})L_{-1}-6706L_{-5}
    \\\noalign{\smallskip} & 
      \ph+7556L_{-3}L_{-2})\Vac{\fr{1}{4};\del_{\theta}a(\theta)}
    \\\noalign{\smallskip} 
  (1,1) &
    (3600L_{-5}+300L_{-1}^5-1200L_{-2}L_{-1}^3-1800L_{-3}L_{-1}^2)
    \Vac{1;a(\theta)}\hfill+(157L_{-1}^5-1100L_{-2}L_{-1}^3
    \\\noalign{\smallskip} & 
      \ph-1242L_{-3}L_{-1}^2+(-512L_{-2}^2+2400L_{-4})L_{-1}+256L_{-3}L_{-2}
      +7540L_{-5})\Vac{1;\del_{\theta}a(\theta)}
    \\\noalign{\smallskip} 
  \end{array}
  $$ 
  $$ 
  \begin{array}{c|l}
  (h,c) & \vac{\chi^{(5)}_{h,c}}\ph\ph{\rm (continued)}
    \\\noalign{\smallskip}\hline\hline\noalign{\smallskip}
  (\frac{9}{4},1) &
    (-525L_{-1}^5+5250L_{-2}L_{-1}^3-2100L_{-3}L_{-1}^2+(-5250L_{-4}
    -4725L_{-2}^2)L_{-1}-9450L_{-3}L_{-2}
    \\\noalign{\smallskip} & 
      \ph+23625L_{-5})\Vac{\fr{9}{4};a(\theta)}
    \\\noalign{\smallskip} & 
    +\,(-646L_{-1}^5+7860L_{-2}L_{-1}^3-6504L_{-3}L_{-1}^2+(-8334L_{-2}^2
    -7860L_{-4})L_{-1}+54270L_{-5}
    \\\noalign{\smallskip} & 
      \ph-16668L_{-3}L_{-2})\Vac{\fr{9}{4};\del_{\theta}a(\theta)}
    \\\noalign{\smallskip} 
  (4,1) &
    (10920L_{-1}^5-218400L_{-2}L_{-1}^3+589680L_{-3}L_{-1}^2+(-1703520L_{-4}
    +698880L_{-2}^2)L_{-1}
    \\\noalign{\smallskip} & 
      \ph-1397760L_{-3}L_{-2}+2271360L_{-5})\Vac{4;a(\theta)}
    \\\noalign{\smallskip} & 
    +\,(22973L_{-1}^5-495860L_{-2}L_{-1}^3+1491702L_{-3}L_{-1}^2
    +(1703232L_{-2}^2-4610268L_{-4})L_{-1}
    \\\noalign{\smallskip} & 
      \ph+6714864L_{-5}-3755904L_{-3}L_{-2})\Vac{4;\del_{\theta}a(\theta)}
    \\\noalign{\smallskip} 
  (-\frac{21}{4},25) &
    (-22770L_{-1}^5-227700L_{-2}L_{-1}^3-1001880L_{-3}L_{-1}^2+(-3051180L_{-4}
    -204930L_{-2}^2)L_{-1}
    \\\noalign{\smallskip} & 
      \ph-409860L_{-3}L_{-2}-4713390L_{-5})\Vac{\fr{-21}{4};a(\theta)}
    \\\noalign{\smallskip} & 
    +\,(-14173L_{-1}^5-81010L_{-2}L_{-1}^3-210716L_{-3}L_{-1}^2+(-18261L_{-2}^2
    -211166L_{-4})L_{-1}
    \\\noalign{\smallskip} & 
      \ph+126477L_{-5}-36522L_{-3}L_{-2})
      \Vac{\fr{-21}{4};\del_{\theta}a(\theta)}
    \\\noalign{\smallskip} 
  (-8,25) &
    (243540L_{-1}^5+4870800L_{-2}L_{-1}^3+27763560L_{-3}L_{-1}^2
    +(119821680L_{-4}+15586560L_{-2}^2)L_{-1}
    \\\noalign{\smallskip} & 
      \ph+62346240L_{-3}L_{-2}+319524480L_{-5})\Vac{-8;a(\theta)}
    \\\noalign{\smallskip} & 
    +\,(-201067L_{-1}^5-4833140L_{-2}L_{-1}^3-30958458L_{-3}L_{-1}^2
    +(-18063808L_{-2}^2
    \\\noalign{\smallskip} & 
      \ph-149094204L_{-4})L_{-1}-437525104L_{-5}
      -80048512L_{-3}L_{-2})\Vac{-8;\del_{\theta}a(\theta)}
    \\\noalign{\smallskip} 
  (-3,25) &
    (494629L_{-1}^5+166300L_{-2}L_{-1}^3-8361514L_{-3}L_{-1}^2
    +(-8372784L_{-2}^2-42795892L_{-4})L_{-1}
    \\\noalign{\smallskip} & 
      \ph-86217192L_{-5}-15363936L_{-3}L_{-2})\Vac{-3;a(\theta)}
    \\\noalign{\smallskip} & 
    +\,(-254016000L_{-1}L_{-4}-31752000L_{-2}L_{-1}^3-111132000L_{-3}L_{-1}^2
    -285768000L_{-5}
    \\\noalign{\smallskip} & 
      \ph-7938000L_{-1}^5)\Vac{-3;\del_{\theta}a(\theta)}
    \\\noalign{\smallskip} & 
    +\,(210105L_{-1}^5+20103300L_{-2}L_{-1}^3+74383470L_{-3}L_{-1}^2
    +(197235360L_{-4}+13547520L_{-2}^2)L_{-1}
    \\\noalign{\smallskip} & 
      \ph+20321280L_{-3}L_{-2}
      +266025060L_{-5})\Vac{-3;\del_{\theta}^2a(\theta)}
    \\\noalign{\smallskip} 
  (-\frac{5}{4},25) &
    (-522018L_{-1}^5-200340L_{-2}L_{-1}^3-1437912L_{-3}L_{-1}^2
    +(-3132108L_{-4}+321678L_{-2}^2)L_{-1}
    \\\noalign{\smallskip} & 
      \ph+449820L_{-3}L_{-2}-2810430L_{-5})\Vac{\fr{-5}{4};a(\theta)}
    \\\noalign{\smallskip} & 
    +\,(112097L_{-1}^5+886810L_{-2}L_{-1}^3+2533260L_{-3}L_{-1}^2
    +(-350215L_{-2}^2+4848726L_{-4})L_{-1}
    \\\noalign{\smallskip} & 
      \ph+4498511L_{-5}-412174L_{-3}L_{-2})
      \Vac{\fr{-5}{4};\del_{\theta}a(\theta)}
    \\\noalign{\smallskip} 
  (-4,26) &
    (-658350L_{-1}^5-4389000L_{-2}L_{-1}^3-15800400L_{-3}L_{-1}^2
    +(-39501000L_{-4}-2633400L_{-2}^2)L_{-1}
    \\\noalign{\smallskip} & 
      \ph-5266800L_{-3}L_{-2}-52668000L_{-5})\Vac{-4;a(\theta)}
    \\\noalign{\smallskip} & 
    +\,(-66567L_{-1}^5+1411570L_{-2}L_{-1}^3+6502092L_{-3}L_{-1}^2
    +(3205032L_{-2}^2+21063180L_{-4})L_{-1}
    \\\noalign{\smallskip} & 
      \ph+37048440L_{-5}+7128264L_{-3}L_{-2})\Vac{-4;\del_{\theta}a(\theta)}
    \\\noalign{\smallskip} 
  (-2,28) &
    (-235620L_{-1}^5-415800L_{-2}L_{-1}^3-1413720L_{-3}L_{-1}^2
    +(-2827440L_{-4}+110880L_{-2}^2)L_{-1}
    \\\noalign{\smallskip} & 
      \ph+110880L_{-3}L_{-2}-2716560L_{-5})\Vac{-2;a(\theta)}
    \\\noalign{\smallskip} & 
    +\,(33601L_{-1}^5+535350L_{-2}L_{-1}^3+1481478L_{-3}L_{-1}^2
    +(93608L_{-2}^2+3042156L_{-4})L_{-1}
    \\\noalign{\smallskip} & 
      \ph+3302084L_{-5}+180728L_{-3}L_{-2})\Vac{-2;\del_{\theta}a(\theta)}
    \\\noalign{\smallskip} 
  (-\frac{27}{8},28) &
    (-554268L_{-1}^5-2771340L_{-2}L_{-1}^3-8868288L_{-3}L_{-1}^2
    +(-19676514L_{-4}-1247103L_{-2}^2)L_{-1}
    \\\noalign{\smallskip} & 
      \ph-2494206L_{-3}L_{-2}-23694957L_{-5})\Vac{\fr{-27}{8};a(\theta)}
    \\\noalign{\smallskip} & 
    +\,(9632L_{-1}^5+1526208L_{-2}L_{-1}^3+5327280L_{-3}L_{-1}^2
    +(2238744L_{-2}^2+13829124L_{-4})L_{-1}
    \\\noalign{\smallskip} & 
      \ph+0919684L_{-5}+4477488L_{-3}L_{-2})
    \Vac{\fr{-27}{8};\del_{\theta}a(\theta)}
    \\\noalign{\smallskip} 
  \end{array}
  $$ 
  $$ 
  \begin{array}{c|l}
  (h,c) & \vac{\chi^{(5)}_{h,c}}\ph\ph{\rm (continued)}
    \\\noalign{\smallskip}\hline\hline\noalign{\smallskip}
  (-5,28) &
    (-24024L_{-1}^5-240240L_{-2}L_{-1}^3-864864L_{-3}L_{-1}^2
    +(-2306304L_{-4}-384384L_{-2}^2)L_{-1}
    \\\noalign{\smallskip} & 
      \ph-960960L_{-3}L_{-2}-3843840L_{-5})\Vac{-5;a(\theta)}
    \\\noalign{\smallskip} & 
    +\,(-5792L_{-1}^5+28595L_{-2}L_{-1}^3+165576L_{-3}L_{-1}^2
    +(269976L_{-2}^2+713808L_{-4})L_{-1}
    \\\noalign{\smallskip} & 
      \ph+2087148L_{-5}+867132L_{-3}L_{-2})\Vac{-5;\del_{\theta}a(\theta)}
    \\\noalign{\smallskip} 
  (-4,33) &
    (-12186720L_{-1}^5-81244800L_{-2}L_{-1}^3-235609920L_{-3}L_{-1}^2
    +(-503717760L_{-4}
    \\\noalign{\smallskip} & 
      \ph-86661120L_{-2}^2)L_{-1}-173322240L_{-3}L_{-2}
      -671623680L_{-5})\Vac{-4;a(\theta)}
    \\\noalign{\smallskip} & 
    +\,(81693L_{-1}^5+37104780L_{-2}L_{-1}^3+121415478L_{-3}L_{-1}^2
    +(98074688L_{-2}^2
    \\\noalign{\smallskip} & 
      \ph+308044644L_{-4})L_{-1}+546134192L_{-5}
      +217814656L_{-3}L_{-2})\Vac{-4;\del_{\theta}a(\theta)}
    \\\noalign{\smallskip} 
  (-\frac{11}{4},33) &
    (-30421755L_{-1}^5-101405850L_{-2}L_{-1}^3-283936380L_{-3}L_{-1}^2
    +(-547591590L_{-4}
    \\\noalign{\smallskip} & 
      \ph-30421755L_{-2}^2)L_{-1}-60843510L_{-3}L_{-2}
      -578013345L_{-5})\Vac{\fr{-11}{4};a(\theta)}
    \\\noalign{\smallskip} & 
    +\,(2904546L_{-1}^5+81047140L_{-2}L_{-1}^3+219449816L_{-3}L_{-1}^2
    +(77929626L_{-2}^2
    \\\noalign{\smallskip} & 
      \ph+457905228L_{-4})L_{-1}+582191814L_{-5}
      +136340532L_{-3}L_{-2})\Vac{\fr{-11}{4};\del_{\theta}a(\theta)}
    \\\noalign{\smallskip} 
  (-\frac{7}{2},\frac{77}{2}) &
    (-66512160L_{-1}^5-332560800L_{-2}L_{-1}^3-848030040L_{-3}L_{-1}^2
    +(-1596291840L_{-4}
    \\\noalign{\smallskip} & 
      \ph-266048640L_{-2}^2)L_{-1}-465585120L_{-3}L_{-2}
      -1862340480L_{-5})\Vac{\fr{-7}{2};a(\theta)}
    \\\noalign{\smallskip} & 
    +\,(4370107L_{-1}^5+196391660L_{-2}L_{-1}^3+526060563L_{-3}L_{-1}^2
    +(360913408L_{-2}^2
    \\\noalign{\smallskip} & 
      \ph+1102564968L_{-4})L_{-1}+1629758776L_{-5}
      +659098684L_{-3}L_{-2})\Vac{\fr{-7}{2};\del_{\theta}a(\theta)}
  \end{array}
  $$}

%
%%< REFERENCES >%%%%%%%%%%%%%%%%%%%%%%%%%%%%%%%%%%%%%%%%%%%%%%%%%%%%%%%%%%%
%
  
  \end{document}